\UseRawInputEncoding
\documentclass[preprint,12pt]{elsarticle}

\usepackage{amssymb}
\usepackage{amsmath}

\usepackage{lineno}

\usepackage{booktabs}
\usepackage{hyperref} 
\usepackage{multirow}
\usepackage{rotating}

\usepackage{tabularx}

\journal{Information and Software Technology}

\begin{document}

\begin{frontmatter}



\title{Domain-Driven Design in Practice: A Large-Scale Empirical Characterisation of the Open-Source Ecosystem}


\author[TUe]{Ozan \"{O}zkan}
\author[TUe,WUR]{\"{O}nder Babur}
\author[TUe]{Mark van den Brand}

\affiliation[TUe]{organization={Mathematics and Computer Science, Eindhoven University of Technology},
 city={Eindhoven},
 country={The Netherlands}}
\affiliation[WUR]{organization={Information Technology Group, Wageningen University \& Research},
 city={Wageningen},
 country={The Netherlands}}

\begin{abstract}
\textbf{Context:} Domain-Driven Design (DDD) has emerged as a dominant paradigm for managing software complexity, yet academic research remains largely confined to theoretical proposals and anecdotal case studies. Our prior research indicates that nearly 39\% of DDD studies lack rigorous empirical evaluation, leaving the practical implementation of the paradigm largely unexamined at scale.

\textbf{Objective:} This study aims to provide the first large-scale characterisation of the DDD landscape within GitHub open-source ecosystem, establishing a data-driven baseline for how the paradigm is implemented and sustained in practice.

\textbf{Method:} We employed a Mining Software Repositories (MSR) methodology, using a hybrid mining strategy (topics and README keywords) to identify an initial set of 11,742 repositories using DDD. To address the peril of ``label noise'', we implemented a novel semantic validation pipeline using GPT-4o with a triplicate majority-vote strategy, yielding a high-fidelity dataset of 2,502 verified repositories. We validated this pipeline against a manually labelled sample and obtained substantial agreement with human experts ($\kappa = 0.77$).

\textbf{Results:} DDD adoption accelerated sharply after an inflection point in 2017, and the resulting projects are notably long-lived as their median lifespan exceeds that of the typical GitHub project by more than an order of magnitude, pointing to sustained, professional-grade engineering rather than short-lived experiments. Layered and Clean Architecture are the dominant structural patterns, while CQRS and Event Sourcing recur in distributed, data-intensive systems. Notably, the data challenge the Java-centric assumption of much academic work: C\# and TypeScript, rather than Java, are the leading languages of practical DDD adoption.

\textbf{Conclusions:} DDD has matured into a stable, professional-grade engineering practice, adopted across a diverse range of languages and application domains. However, a quarter of the projects (25.3\%) record no explicit business context in their documentation, revealing a persistent gap between how domain intent is designed and how it is preserved in version-controlled artifacts. We argue for lightweight architectural traceability standards to close this gap, and we provide practical guidance for teams that reuse these repositories as reference implementations.
\end{abstract}

\begin{graphicalabstract}
\end{graphicalabstract}

\begin{highlights}
\item We characterise DDD across 2,502 verified open-source GitHub repositories
\item An agentic GPT-4o pipeline reaches Cohen's k=0.77 against expert labelling
\item DDD adoption accelerates sharply after a 2017 inflection point
\item C\# (34\%) and TypeScript (18\%) lead practical DDD, not Java (16\%)
\item 25.3\% of DDD projects record no explicit business domain in metadata
\end{highlights}


\end{frontmatter}


 
\section{Introduction} \label{sec:introduction}

The growing complexity of modern software systems has become one of the defining challenges of contemporary software engineering. Enterprise applications today routinely span thousands of interconnected components, serve millions of concurrent users, and encode intricate business rules that evolve continuously alongside the organisations they support \cite{evans_domain-driven_2004}. Empirical studies of software evolution show that the majority of source code lines are modified or deleted within their first two years of existence, reflecting the continuous churn imposed by evolving business requirements \cite{spinellis_software_2021}. Unlike the physical constraints of traditional engineering disciplines, software complexity is largely invisible, accumulating silently in codebases until it manifests as defects, missed deadlines, or systems that resist change. 
Managing this complexity, not merely at the technical level but at the boundary between technology and the business domain it serves, has therefore become a central concern for both researchers and practitioners \cite{evans_domain-driven_2004, ozkan_domain-driven_2025}.

Among these approaches, \textit{Domain-Driven Design (DDD)} has emerged as a dominant paradigm for solving software architecture problems through the core tenets of Ubiquitous Language, Bounded Contexts, and Rich Domain Models \cite{evans_domain-driven_2004, vernon_implementing_2013}. Pioneered by Eric Evans in 2004, DDD seeks to bridge the gap between technical jargon and domain-specific vocabulary, ensuring that the software reflects the real-world business requirements \cite{evans_domain-driven_2004, ozkan_refactoring_2023-1}. While the paradigm has enjoyed significant industrial advocacy, particularly within the .NET and Java communities, academic research has remained largely siloed within theoretical design proposals and anecdotal case studies \cite{ozkan_domain-driven_2025}. Our prior Systematic Literature Review (SLR) revealed that approximately 39\% of DDD studies lack rigorous empirical evaluation, relying instead on subjective observations rather than data-driven baselines \cite{ozkan_domain-driven_2025}.

To address this gap, our study applies the \textit{Mining Software Repositories (MSR)} methodology to provide the first large-scale characterization of the DDD landscape on GitHub. How DDD is actually implemented at scale remains poorly understood.

In this study, we mined and semantically validated 2,502 repositories on GitHub using DDD, treating them as records of real-world software development activity \cite{guemes_emerging_2018}. Here, semantic validation refers to verifying that a repository genuinely implements DDD at the code and structural level, rather than merely referencing the term in its metadata or documentation. To address the documented ``perils'' of label noise and reproducibility \cite{kalliamvakou_promises_2016, tutko_how_2022}, we developed a novel semantic validation pipeline based on LLMs \cite{hou_large_2023} that distinguishes genuine ``Rich Domain Models'' from anemic implementations at scale. Crucially, we did not accept these labels on trust: we benchmarked the pipeline against an independently hand-labelled sample of 50 repositories, where it reached substantial agreement with human experts (Cohen's $\kappa = 0.77$ and an F1-score of 92.5\%; see Section~\ref{subsec:reliability}).

Our findings uncover a \textit{critical turning point in 2017}, marking the transition of DDD from theoretical infancy to practical maturity in open source \cite{ozkan_domain-driven_2025}. We demonstrate that the DDD ecosystem is composed of \textit{professional-grade engineered systems}, with a median project longevity of 340.37 days (mean: 660.73 days), more than 34 times the 9.9-day median of the typical GitHub project \cite{kalliamvakou_promises_2016}. Furthermore, our data challenges the Java-centric academic tradition by identifying C\# (34.17\%) and TypeScript (17.71\%) as the primary drivers of industrial adoption. Ultimately, this research provides the quantitative baseline necessary to move the architectural discourse from opinion-based argument toward data-driven evidence.

The remainder of this paper is organized as follows: Section~\ref{sec:related_work} situates our work within the DDD and MSR literature; Section~\ref{sec:research_objectives_and_method} describes our AI-assisted mining methodology; Section~\ref{sec:results} presents our empirical findings; Section~\ref{sec:discussions} and~\ref{sec:implications} synthesize the results for researchers and practitioners; Section~\ref{sec:limitations_threats_to_validity} addresses threats to validity; and Section~\ref{sec:conclusion} concludes the study.

\subsection{Domain-Driven Design Essentials} \label{subsec:domain_driven_design}
DDD is an approach to software development that aims to keep the structure of a system's code in step with the complexity of the business domain it addresses. Its purpose is to narrow the distance between technical implementation and business needs, so that developers and domain experts can communicate more effectively \cite{evans_domain-driven_2004, vernon_implementing_2013}. The paradigm originates with Eric Evans, whose influential book \textit{Domain-Driven Design: Tackling Complexity in the Heart of Software} \cite{evans_domain-driven_2004} established its foundations. Evans contends that building software well demands a thorough grasp of the domain's subtleties, so that the resulting system genuinely reflects real-world needs rather than a superficial approximation of them \cite{evans_domain-driven_2004}.
 
Within DDD, the \textit{domain} denotes the particular field or subject the application addresses, whether e-commerce, finance, healthcare, or logistics. The approach centres on building a rich, well-defined domain model that faithfully captures the rules, processes, and entities of that field. Such a model is produced collaboratively by developers and domain experts, typically through a shared vocabulary, the \textit{Ubiquitous Language}, that unifies technical and domain terminology.
 
The concepts introduced in the following subsections give software architects a principled way to represent these real-world complexities. Applied together, they let engineers reconcile technical precision with a deep understanding of the domain.
 
\subsubsection{Ubiquitous Language}
Ubiquitous Language is foundational to DDD: it is the practice of establishing a single shared vocabulary that closes the gap between technical jargon and the language of the domain \cite{evans_domain-driven_2004}. By giving developers, domain experts, and business stakeholders common terms, it removes ambiguity from their discussions and keeps communication consistent. Because this vocabulary is carried directly into the code, domain knowledge is translated into the system's design and implementation without distortion.
 
\subsubsection{Bounded Context}
A Bounded Context delimits the zone within which the terms of a particular domain model carry a single, consistent meaning. In large applications that span several domain models, terminology can easily become ambiguous or inconsistent; Bounded Contexts address this by separating those models so that each retains its integrity \cite{evans_domain-driven_2004}. Each context confines a model to a well-defined scope, preventing semantic clashes and giving the business domain a coherent representation within its boundary. A context may also stand for a (sub-)system or a team, aligning model boundaries with software components, organisational structure, and responsibilities.
 
\subsubsection{Context Mapping} \label{subsubsec:context_mapping}
Context Mapping is the strategic practice of understanding and governing the relationships among the different bounded contexts of a complex system \cite{evans_domain-driven_2004}. Although a Bounded Context fixes the scope in which a model applies, real systems usually involve several teams working on distinct subdomains, each with its own context. Those contexts must still interoperate, which raises the risk that one model will contaminate or be misread by another.
 
To manage this, Context Mapping offers a visual and conceptual means of stating explicitly how bounded contexts relate and how their integration should be governed. It guides teams toward suitable integration patterns, such as Shared Kernel, Customer/Supplier, Conformist, or Anti-Corruption Layer, according to how the teams collaborate and what legacy or organisational constraints apply. The resulting Context Map becomes a key communication artefact in large, multi-team systems.
 
Evans stresses that strategic design choices matter most when teams build models in parallel, and that context mapping acts as the bridge that keeps those models aligned and free of conflict \cite{evans_domain-driven_2004}. In practice, context maps also shape technical decisions about API design, data ownership, and service integration, making them central to preserving semantic consistency and architectural integrity across distributed systems.
 
\subsubsection{Aggregates}
An Aggregate groups Entities and Value Objects together with the domain logic expressed through entity methods and domain services. Aggregates act as the controlled entry points for reading and modifying data, so that business transactions preserve data integrity \cite{evans_domain-driven_2004}. By keeping data within explicit boundaries, they improve data management and scalability while enforcing the logical limits that prevent uncontrolled modification.
 
\subsubsection{Entities and Value Objects}
DDD draws a deliberate distinction between Entities and Value Objects. An Entity represents a concrete element of the domain that carries a unique identity and changes over time, capturing the mutable side of the business reality \cite{evans_domain-driven_2004}. A Value Object, by contrast, describes a conceptual attribute that has no identity and is immutable. Separating the two lets DDD model both tangible, identity-bearing things and abstract attributes, yielding a more complete picture of the domain \cite{evans_domain-driven_2004}.
 
\subsubsection{Domain Services}
Domain Services hold domain logic that does not belong to any single Entity or Value Object \cite{evans_domain-driven_2004}. They coordinate operations and interactions that span several objects, and by isolating this cross-cutting logic they keep the code modular, reusable, and focused. In effect, they mirror the way coordinated processes play out in the real world, tightening the fit between the software and the domain.
 
\subsubsection{Domain Events}
A Domain Event models a noteworthy change or occurrence within the domain and communicates it to the rest of the system \cite{vernon_implementing_2013}. Such an event is a lightweight object describing something significant that has already taken place, and is conventionally named in the past tense to signal this. Domain Events decouple the parts of a system: components or aggregates can react to one another without depending on each other's internals, which supports a more modular and maintainable architecture.
 
\subsubsection{Anti-Corruption Layer (ACL)}
An Anti-Corruption Layer shields a domain model from outside influences by mediating its interactions with other models or systems. Translating data between the differing representations on each side, the ACL keeps the domain model consistent and protects it from external distortion \cite{evans_domain-driven_2004}.
 
\subsubsection{Core Domain}
The Core Domain is a strategic-design concept denoting the part of a system that creates the most value and sets the business apart from its competitors. Because it is where competitive advantage is won, development effort is concentrated there and the strongest developers are typically assigned to it \cite{evans_domain-driven_2004}.
 
\subsubsection{Shared Kernel}
A Shared Kernel is a pattern in which two Bounded Contexts deliberately share a common portion of the domain model. That shared portion has to be managed carefully to avoid conflicts and to keep the model coherent, which usually calls for close coordination between the teams involved \cite{evans_domain-driven_2004}.
 
The ACL has parallels in several established design patterns. The \textit{Remote Proxy} \cite{gamma_design_2009} supplies a local stand-in for an object that lives in another address space, brokering interaction between local and remote systems. The \textit{Wrapper} \cite{buschmann_pattern-oriented_1996} wraps an object to expose a different interface, much as the ACL wraps external systems behind a consistent one. The \textit{Adapter} \cite{gamma_design_2009} reshapes a class's interface into the one a client expects, just as the ACL reshapes external data representations to suit the internal domain model.
 
The essential value of DDD lies in producing software whose structure closely tracks the way the business itself reasons about its domain \cite{evans_domain-driven_2004, vernon_implementing_2013}. By bringing domain experts, business stakeholders, and developers together, DDD enables the kind of collaboration that yields solutions well matched to business needs, deepens shared understanding of the problems and their solutions across the team, and strengthens working relationships between teams \cite{evans_domain-driven_2004, brandolini_eventstorming_2013}.
 
\subsubsection{Strategic and Tactical DDD}
The DDD community commonly organises these patterns into two groups, \textit{Strategic} and \textit{Tactical} design, a framing rooted in Evans \cite{evans_domain-driven_2004} and popularised in practitioner literature such as Vernon's \cite{vernon_implementing_2013}. Strategic DDD is concerned with aligning the software model with the wider business strategy and organisation, and covers concepts such as Bounded Context, Core Domain, Context Maps, and the Anti-Corruption Layer. Tactical DDD, in contrast, comprises the concrete modelling techniques applied inside a single Bounded Context, including Entities, Value Objects, Aggregates, Domain Services, and Domain Events. We adopt this distinction between strategic and tactical design throughout the study as an analytical lens for interpreting how DDD concepts are applied across the repositories we examine.
 
\subsubsection{Event Storming}
Beyond its modelling patterns, DDD also offers collaborative modelling techniques that help connect business and technical stakeholders. The best known is Event Storming, a workshop-based method introduced by Alberto Brandolini \cite{brandolini_eventstorming_2013} in which domain experts and developers jointly explore business processes by surfacing and arranging domain events, commands, aggregates, and actors. It is especially useful for uncovering Bounded Contexts and for aligning the domain model with the way the business actually operates \cite{uludag_supporting_2018}.
 
\subsubsection{Other DDD Patterns and Concepts}
The patterns and concepts covered above are those that featured most prominently in our earlier systematic literature review \cite{ozkan_domain-driven_2025}. The DDD literature contains further patterns, among them Customer/Supplier, Open Host Service, Conformist, and Published Language, but our aim here is a focused account of the patterns most central to that review rather than an exhaustive catalogue of every DDD concept.

To make these concepts concrete, consider a small insurance example. Under DDD, the Ubiquitous Language surfaces directly in the code: a \textit{Policy} is an Entity with its own identity and lifecycle, \textit{Premium} and \textit{Coverage} are immutable Value Objects, and a \textit{Claim} is an Aggregate that enforces the rules governing how a claim moves from \textit{Filed} to \textit{UnderReview} to \textit{Settled}. Business invariants, such as the rule that a claim cannot be settled before it has been reviewed, reside inside the \textit{Claim} aggregate rather than in external procedural code; this is what is meant by a ``Rich Domain Model''. An \textit{anemic} design would instead store the same data on a passive \textit{ClaimRecord} and scatter the rules across service classes and database triggers, so that the code no longer reflects how the business itself reasons about claims.

\section{Related Work} \label{sec:related_work}
Research into DDD has seen a steady increase in interest, particularly since 2017 \cite{ozkan_domain-driven_2025}. However, our prior SLR of 36 peer-reviewed studies revealed that the landscape is predominantly composed of theoretical proposals, with 39\% lacking rigorous empirical evaluation \cite{ozkan_domain-driven_2025}. Existing empirical studies are largely confined to small-scale case studies or action research in single industrial contexts \cite{ozkan_refactoring_2023-1, uludag_supporting_2018, krause_microservice_2020, maddodi_aggregate_2020}, leaving the large-scale practical adoption of DDD across diverse ecosystems largely unexplored. This reflects a broader ``culture gap'' identified by Robles et al. \cite{robles_reflection_2023}, who noted that the intersection between the repository mining and architectural modeling communities remains shallow, with minimal research overlap.

Our work is methodologically situated within the tradition of domain-specific landscape characterisations on GitHub. This tradition builds on earlier large-scale efforts to mine software modeling artifacts, such as Hebig et al.'s \cite{hebig_quest_2016} study of UML usage across open-source projects. Kochanthara et al. \cite{kochanthara_painting_2022, cosentino_systematic_2017} and Rigas et al. \cite{rigas_mining_2023} successfully mapped the Automotive and Electric Vehicle sectors respectively, demonstrating how MSR techniques can reveal technological shifts that are often invisible in general-purpose datasets. Most recently, Saeedi Nikoo et al. \cite{nikoo_empirical_2025} characterised the BPMN landscape, establishing a 16-domain taxonomy whose categories they aligned with the GICS \cite{msci_gics_2024} and TRBC \cite{refinitiv_trbc_2020} industry classification schemes, highlighting the industrial diversity of modeling artifacts. Our study directly extends this tradition to the DDD paradigm, which unlike automotive software or BPMN has not previously been characterised at scale.

A key methodological contribution of our work is the use of LLMs for semantic validation at scale. Hou et al. \cite{hou_large_2023} provide a comprehensive survey of LLM applications in software engineering, identifying classification, code analysis, and architectural reasoning as emerging strengths. While prior MSR studies have relied on keyword matching or topic modeling for repository classification \cite{hindle_automated_2011, panichella_how_2013}, these approaches are known to produce high label noise when applied to architectural concepts \cite{tutko_how_2022}. Our agentic GPT-4o pipeline addresses this limitation by iteratively inspecting source code artifacts to confirm structural DDD intent, rather than relying solely on metadata signals; we describe it in full in Section~\ref{subsec:semantic_classification}. To our knowledge, this is the first application of an agentic LLM validation strategy in a DDD MSR landscape study, and we evaluate its reliability and validity against a manually labelled benchmark in Section~\ref{subsec:reliability}.

While these studies provide essential baselines, Tutko et al. \cite{tutko_how_2022} highlight a persistent reproducibility crisis in MSR, where 33\% of studies lack sufficient retrieval documentation. Furthermore, AlMarzouq et al. \cite{almarzouq_mining_2020} argue that the academic focus on Java-based projects often masks the practical realities of industrial adoption. Our study addresses both concerns: we provide full retrieval documentation including per-query noise rates, and our results reveal a technologically diverse ecosystem where C\# and TypeScript dominate over Java, directly challenging the Java-centric bias documented in the literature.

\section{Research Objectives and Method} \label{sec:research_objectives_and_method}
Building on our previous SLR, which identified a scarcity of empirical evaluation in the DDD literature \cite{ozkan_domain-driven_2025}, this study applies a MSR methodology to establish a data-driven baseline for the domain.

We utilised the GitHub GraphQL API to mine an initial dataset of 11,742 repositories, treating them as verifiable records of development activity \cite{guemes_emerging_2018}. Our repository identification process follows the tradition of domain-specific landscape studies \cite{kochanthara_painting_2022, rigas_mining_2023}, employing a hybrid strategy of topic-based mining and README keyword analysis. This dual approach allows us to capture both technical intent and business context. To guarantee the robustness of our dataset, we applied rigorous filtering criteria derived from established MSR guidelines and peril-avoidance recommendations \cite{kalliamvakou_promises_2016,ralph_empirical_2020}, specifically targeting original, non-forked repositories with sustained activity levels.

\subsection{Goal and Research Questions} \label{subsec:goal_and_research_questions}
The primary goal of this study is to empirically characterise the state of DDD within the open-source community. Unlike previous theoretical studies, we aim to provide a quantitative baseline of how DDD adoption has manifested and evolved in public repositories. To achieve this, we formulate the following main research question:

\begin{itemize}
 \item \textbf{RQ\textsubscript{M}:} What are the characteristics and trends of DDD adoption in open-source projects on GitHub?
\end{itemize}
To provide a comprehensive answer, we decompose this main question into descriptive sub-questions (RQs):

\begin{itemize}
 \item \textbf{RQ\textsubscript{1} (Temporal Evolution):} How has the prevalence and activity of DDD-related repositories on GitHub evolved since its earliest appearance in open source, and is the interest sustained?
 \item \textbf{RQ\textsubscript{2} (Architectural Taxonomy):} What is the distribution of high-level architectural styles (e.g., Clean, Onion, Hexagonal, or Layered Architecture) within the verified DDD ecosystem?
 \item \textbf{RQ\textsubscript{3} (Exemplary Projects):} What are the characteristics of the most technically intensive DDD projects identified by the volume of sustained commit activity, and what business domains and architectural patterns do they showcase?
 \item \textbf{RQ\textsubscript{4} (Ownership and Stakeholders):} What is the distribution of repository ownership (Individual vs. Organisation) and how does it influence the rigour of DDD adoption?
 \item \textbf{RQ\textsubscript{5} (Technological and Business Ecosystems):} Which programming languages and application business domains (e.g., Finance, e-Commerce, Healthcare) dominate the DDD landscape?
 \item \textbf{RQ\textsubscript{6} (Community Engagement and Sustainability):} How does the community interact with DDD projects in terms of popularity and maintenance dynamics, such as Pull Request (PR) latency and activity frequency?
\end{itemize}

\subsection{Data Sources} \label{subsec:search_strategy_and_data_sources}
In this study, we used GitHub to collect DDD software repositories and descriptive information from the repositories collected. Our motivation for using GitHub for this process is the size of the open-source software repositories hosted on the platform and its industry-wide recognition. As of 2025, there are 420+ million repositories hosted and 150+ million developers registered on GitHub, alongside 4+ million organisations~\cite{github_about_2025}.

\begin{figure}
 \includegraphics[width=\textwidth]{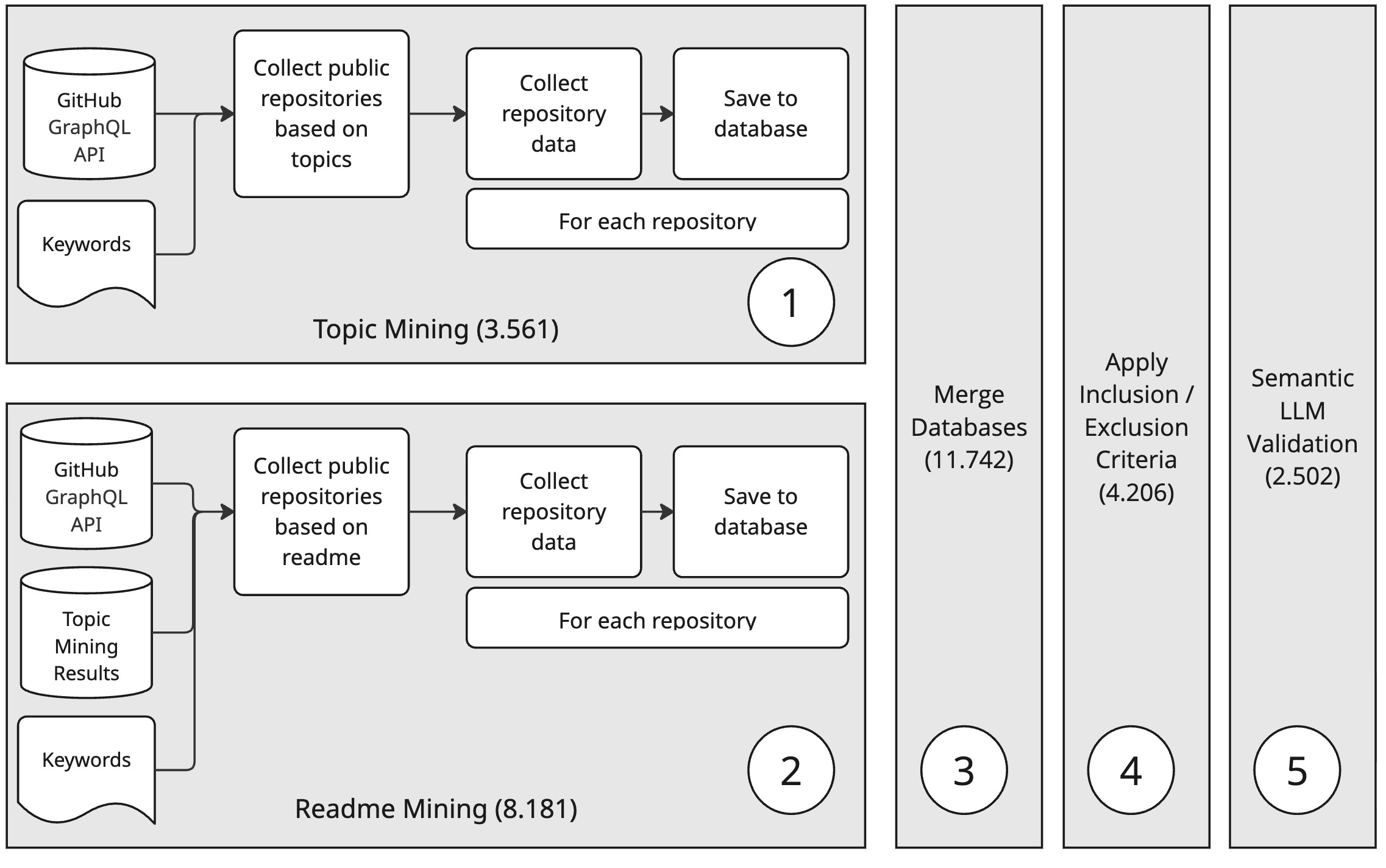}
\caption{Overview of the data collection and verification pipeline, comprising topic-based mining (Step 1; 3,561 repositories), README-based mining (Step 2; 8,181 repositories), database merging (Step 3; 11,742), inclusion/exclusion filtering (Step 4; 4,206), and semantic LLM validation (Step 5), resulting in 2,502 validated repositories.}
\label{fig:data_collection}
\end{figure}

Figure~\ref{fig:data_collection} provides an overview of our data collection and verification pipeline. Our data collection process involved using the GitHub GraphQL API, which enabled structured and efficient retrieval of up-to-date information from publicly available repositories. For each repository, we collected detailed metadata including its name, owner information, creation and last update timestamps, description, default branch, primary programming language, and indicators of project activity such as the number of stars, forks, watchers, open issues, and contributors. In addition, we gathered file-level data from the repository's default branch to capture the composition of the source code and documentation. The README contents and topic labels were also extracted to understand each project’s self-description and conceptual associations. 

To capture development dynamics, we collected commit data containing commit identifiers, authorship information, timestamps, and associated code changes (insertions, deletions, and files modified). Likewise, we extracted detailed information on issues, pull requests, and their associated comments, including creation, update, and closure dates, comment authors, and discussion content. This allowed us to analyze both technical and social aspects of repository activity. Complementing the API-based data, each repository was also cloned from its default branch to verify data consistency and to analyze source files more thoroughly, ensuring that the most recent version of the codebase was examined.

To facilitate the data collection process, we developed a \textit{Python} tool\footnote{https://github.com/OzanOzkan/ddd-landscape-github}. This tool was designed to query specific topics and repositories on GitHub, gather pertinent data, and store this information in a SQLite database. The use of a SQLite database was instrumental in enabling us to create a snapshot of the collected data. Furthermore, it provided us with the capability to perform data analysis using SQL queries and the implementation of simple algorithms. This approach allowed us to systematically analyze the repositories and extract meaningful metrics related to the application of DDD in open-source projects.

\subsection{Primary Repository Identification} \label{subsec:repository_identification_and_selection}
Identifying a specific type of software repository on GitHub requires careful consideration of how projects are labeled and described. One of the most common approaches in prior research has been to use topic modeling or keyword-based classification~\cite{hindle_automated_2011, panichella_how_2013}. However, such methods have been shown to be inefficient and prone to noise when applied to large-scale repository mining~\cite{liew_using_2022}. Instead, we adopted GitHub’s built-in \textit{topics} feature\footnote{\url{https://github.blog/2017-01-31-introducing-topics}}, which allows repository authors or GitHub itself to assign descriptive tags to repositories. This feature has been effectively used in prior mining studies, for example in identifying automotive software repositories for painting the picture of automotive software in open source community~\cite{kochanthara_painting_2022, cosentino_systematic_2017}.

To identify DDD repositories, we combined topic-based and README-based discovery methods. This combined approach ensures both precision and coverage, as prior work has shown that relying solely on GitHub topics can lead to missing relevant projects that do not use topic labeling consistently~\cite{kochanthara_painting_2022, cosentino_systematic_2017}. README-based search, on the other hand, can reveal repositories that describe DDD concepts textually even if no topic labels are applied. Similar hybrid strategies have been successfully employed in studies mining open-source projects in other domains~\cite{kochanthara_painting_2022, cosentino_systematic_2017, chabert_experienced_2022} to obtain more comprehensive and robust datasets.

The discovery process covered all repositories created up to end of 2025 and relied on a curated list of DDD-specific keywords derived from the seminal work of Evans~\cite{evans_domain-driven_2004} and our systematic literature review \cite{ozkan_domain-driven_2025}. The final keyword set, organized by conceptual scope, is shown in Table~\ref{tab:ddd_keywords}. Using these keywords, topic-based mining (Step~1) returned 3{,}561 public repositories, while README-based mining (Step~2) identified an additional 8{,}181 repositories referencing DDD concepts in their documentation. Merging these two sets yielded a combined pool of 11{,}742 candidate repositories (Step~3). 

Together, these searches produced an initial candidate pool of 11{,}742 repositories for subsequent filtering and analysis. 

However, relying on GitHub topics and README keywords for repository discovery carries known limitations. Topic labels are author-assigned and inconsistently applied; many repositories implementing DDD concepts do not use any of the keywords in our search set, introducing recall bias \cite{kochanthara_painting_2022, cosentino_systematic_2017}. Conversely, keywords such as \texttt{ddd} are highly ambiguous in free-text contexts, introducing label noise \cite{kalliamvakou_promises_2016, tutko_how_2022}. Furthermore, GitHub's search API does not guarantee exhaustive retrieval. Results are ranked by relevance rather than completeness, meaning that some relevant repositories may not appear in the returned result sets \cite{kalliamvakou_promises_2016}. These limitations motivate the hybrid discovery strategy described above and, critically, the semantic validation pipeline described in Section~\ref{subsec:semantic_classification}, which addresses label noise through LLM-based verification rather than relying solely on keyword presence.

\begin{table}[htbp]
\centering
\begin{tabular}{ll}
\hline
\textbf{Category} & \textbf{Keyword} \\ \hline
Core term & \texttt{domain-driven-design} \\
Abbreviation & \texttt{ddd} \\
Variant term & \texttt{domain-driven-development} \\
Strategic concept & \texttt{bounded-context} \\
Strategic concept & \texttt{ubiquitous-language} \\
Tactical pattern & \texttt{aggregate-root} \\
Tactical pattern & \texttt{value-object} \\
Tactical pattern & \texttt{domain-event} \\
Tactical pattern & \texttt{domain-service} \\
Strategic pattern & \texttt{context-map} \\
Strategic pattern & \texttt{context-mapping} \\
Integration pattern & \texttt{anticorruption-layer} \\ \hline
\end{tabular}
\caption{Keywords used for identifying DDD repositories on GitHub.}
\label{tab:ddd_keywords}
\end{table}

\subsection{Discovery Source Analysis and Noise Characterisation} \label{subsec:discovery_source_analysis}

A known limitation of hybrid keyword discovery is that different query types carry different levels of semantic precision. Topic-based queries rely on author-assigned labels which carry stronger signal, while README-based queries match free text which is inherently more ambiguous \cite{tutko_how_2022,kochanthara_painting_2022}. To quantify this precision difference and motivate the need for semantic validation, we analysed the noise rate of each query source independently by cross-referencing the initial candidate pool with the final verified set. The results of this analysis are reported in Section~\ref{sec:results}.

\subsection{Inclusion and Exclusion Criteria} \label{subsec:inclusion_and_exclusion_criteria}
After collecting the initial dataset of candidate repositories, we applied a series of inclusion and exclusion criteria to ensure that only active (showing measurable development and collaboration), software-relevant (built in general-purpose programming languages rather than documentation or configuration), and complete (original, non-forked, with fully retrievable file contents) projects were retained for analysis. These criteria were derived from established guidelines for mining software repositories~\cite{kalliamvakou_promises_2016, maia_systematic_2022} and applied through structured SQL queries on our normalized SQLite database. The criteria aimed to remove duplicates, forks, and non-software repositories while retaining projects that demonstrate measurable development and collaboration activity. Table~\ref{tab:selection_criteria} provides a summary of the inclusion and exclusion rules.

Repositories were included if they were publicly accessible and represented active software projects with observable development and collaboration. Specifically, we required each repository to be non-archived, non-forked, and created before the end of 2025 to maintain a consistent observation window. In addition, repositories were required to have at least ten commits, at least one issue or pull request, and a non-zero number of lines of code (LOC) to ensure substantial software content and activity. To guarantee that the dataset focuses on code-based systems rather than documentation or tutorials, we restricted inclusion to repositories implemented in general-purpose programming languages commonly used for DDD, such as Java, C\#, Kotlin, Python, and Go \cite{ozkan_domain-driven_2025}.

Repositories were excluded if they were forks, archived, or contained incomplete data due to missing file contents. We also excluded repositories dominated by markup, configuration, or documentation languages, as these are not indicative of executable software systems. The included general-purpose languages were Java, C\#, TypeScript, Kotlin, PHP, JavaScript, Groovy, C++, Rust, Python, Scala, Dart, Go, Ruby, Swift, Elixir, Clojure, F\#, Objective-C, Haskell, Erlang, Crystal, D, Lua, Racket, Nim, Ada, CAP~CDS, Haxe, Rascal, Pony, and Prolog. Excluded languages comprised markup, configuration, and documentation formats (e.g., HTML, CSS, Markdown, XML, YAML, Dockerfile, Makefile, and templating languages), scripting and shell languages (e.g., Shell, PowerShell, Batchfile), notebook and statistical formats (e.g., Jupyter~Notebook, R, Mathematica), query languages (e.g., PLpgSQL, TSQL), and other formats not representing executable software systems. 

Following the application of these criteria (Step~4), a total of 4{,}206 repositories were retained as candidate projects. However, metadata filtering alone cannot confirm the actual implementation of DDD patterns. To address potential label noise, this candidate set was subjected to a deeper semantic validation process (Step~5).

\begin{table}[htbp]
\centering
\begin{tabularx}{\textwidth}{p{0.8cm}p{4.5cm}X}
\hline
\textbf{ID} & \textbf{Criterion} & \textbf{Rationale} \\ \hline
\multicolumn{3}{l}{\textit{Inclusion Criteria}} \\ \hline
I1 & Public repository & Ensures data accessibility and reproducibility through the GitHub API. \\
I2 & Not archived or forked & Removes inactive or duplicate repositories to retain original, active projects. \\
I3 & Created in 2025 and before & Defines a consistent temporal scope for the analysis. \\
I4 & $\geq$10 commits and at least one issue or pull request & Ensures substantial development history and evidence of collaboration or user interaction. \\
I5 & Non-zero lines of code (LOC) & Confirms the presence of actual source code. \\
I6 & General-purpose programming language & Focuses on executable software systems rather than documentation, tutorials, or configuration repositories. \\ \hline
\multicolumn{3}{l}{\textit{Exclusion Criteria}} \\ \hline
E1 & Missing file contents or incomplete crawl & Ensures completeness and data reliability. \\
E2 & Markup, configuration, or documentation language & Removes repositories dominated by non-executable content (e.g., HTML, Markdown, YAML, Dockerfile) that do not represent software systems. \\ \hline
\end{tabularx}
\caption{Inclusion and exclusion criteria applied to construct the study dataset. The GPT-4o classification pipeline (Step 5) inspected source files only for the seven most prevalent extensions (\texttt{.java}, \texttt{.cs}, \texttt{.ts}, \texttt{.js}, \texttt{.py}, \texttt{.php}, \texttt{.go}), representing 95.29\% of repositories by primary language; repositories in other included languages were assessed from metadata, README content, and directory structure.}
\label{tab:selection_criteria}
\end{table}

\subsection{Semantic Classification at Scale via Large Language Models (LLM)} \label{subsec:semantic_classification}
\begin{figure}[htbp]
 \centering
 \includegraphics[width=\textwidth]{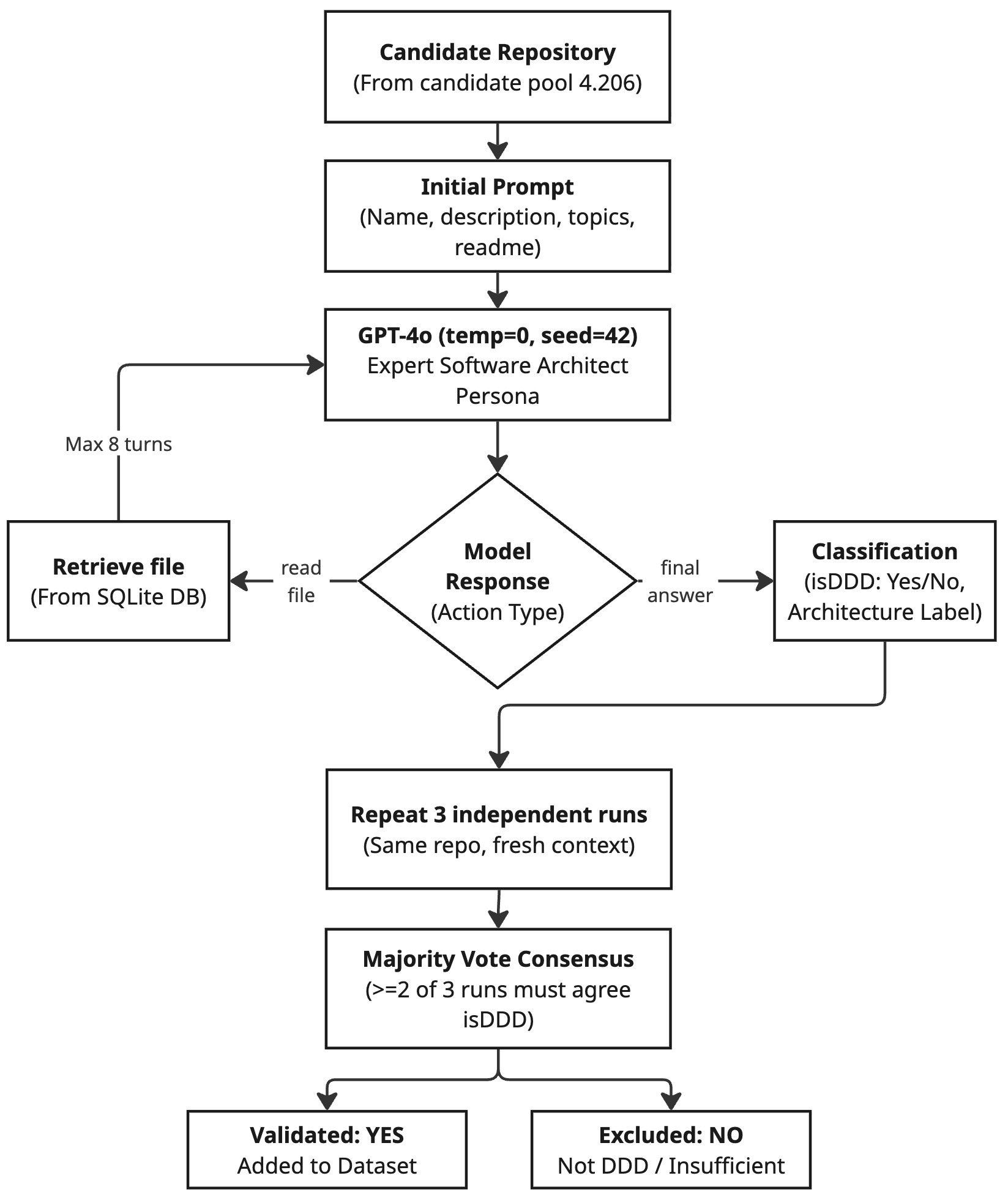}
 \caption{The agentic GPT-4o classification pipeline. Each repository undergoes up to eight conversation turns in which the model may iteratively request and inspect source file contents before emitting a final classification verdict. The process is repeated three times independently per repository; a majority-vote consensus of at least two out of three runs is required for inclusion in the validated dataset.}
 \label{fig:llm_workflow}
\end{figure}

While keyword and topic-based filtering provide a necessary initial scope, they are prone to false positives, such as repositories that tag "DDD" but do not implement its architectural patterns (e.g., sample apps, tutorials, or mislabeled projects). To resolve this, we implemented a semantic validation pipeline using a Large Language Model (GPT-4o) to simulate expert code review at scale (Figure~\ref{fig:llm_workflow}).

We developed a Python-based pipeline to interact with the GPT-4o model deployed via the Azure OpenAI API (\texttt{2024-05-01-preview}). To maximise reproducibility and minimise non-deterministic variance, the \texttt{temperature} parameter was fixed at 0 and the random \texttt{seed} at 42 across all classification runs.

The classification workflow employed an \textit{agentic, iterative context-building strategy} rather than a single-pass prompt. As shown in Figure~\ref{fig:llm_workflow}, it unfolds in three stages: (i)~iterative, multi-turn inspection of repository contents; (ii)~a dual-dimension classification verdict (the \textit{isDDD} label and the architectural style); and (iii)~triplicate execution with majority-vote consensus. We describe each in turn. \textit{Iterative context building.} Each repository was processed through a structured multi-turn conversation with the model. In the first turn, the model received the repository name, description, GitHub topic labels, a preprocessed README snippet (HTML-stripped and stopword-filtered), and a filtered list of up to 300 source code file paths (restricted to \texttt{.java}, \texttt{.cs}, \texttt{.ts}, \texttt{.js}, \texttt{.py}, \texttt{.php}, and \texttt{.go}, the seven most prevalent language extensions in the verified dataset, collectively representing 95.29\% of repositories by primary language), which the model used as a structural map to identify architecturally relevant files for further inspection. These paths were provided in database storage order (the order in which files were retrieved during collection); no relevance-, size-, or name-based ordering was applied. This filter alone reduced the file space considerably. The average repository contained 238 files after filtering, and 82.58\% of repositories fell entirely within the 300-path limit. The cap was therefore reached only for the remaining 16.7\% of repositories, representing larger, more complex codebases. For these cases, only the first 300 filtered paths as indexed in the database were visible to the model, which we acknowledge as a limitation for larger repositories where architecturally relevant files may appear deeper in the directory hierarchy.

\textit{Classification verdict.} The system prompt instructed the model to act as an expert software architect and to classify each repository along two dimensions simultaneously: (1) a binary \textit{isDDD} label (YES/NO), based on evidence of structural DDD patterns such as domain layer isolation, Ubiquitous Language in naming conventions, and the 
presence of tactical building blocks (Aggregates, Value Objects, Repository interfaces); and (2) a dominant \textit{architectural style} label selected from a predefined taxonomy of seven styles common in DDD literature: Layered Architecture, Clean Architecture, Hexagonal Architecture, Onion Architecture, 
CQRS, Event-Driven Architecture, and Microservices. The model was explicitly instructed to apply a liberal threshold for DDD confirmation, that is, to label a project YES if clear structural \textit{intent} was observable, even if the implementation was not theoretically complete. This design choice reflects the 
established distinction between ``Rich Domain Models'' and anemic implementations \cite{evans_domain-driven_2004, ozkan_domain-driven_2025}. 

\textit{Triplicate majority vote.} To further mitigate the non-deterministic nature of generative models, we employed a \textit{Triplicate Validation Strategy}. Each repository was classified three times independently. A repository was only included in the final dataset if at least two out of the three runs confirmed the presence of DDD patterns (majority-vote consensus). This rigorous semantic filter identified and excluded 1{,}704 repositories that lacked sufficient architectural evidence, reducing the dataset from 4{,}206 candidates to a final, high-fidelity set of \textbf{2{,}502 confirmed repositories}.

\subsubsection{Architectural Style Classification}
\label{subsubsec:architectural_classification}

In addition to the binary DDD validation, each repository was simultaneously classified into a dominant architectural style during the same iterative GPT-4o pass. The model was instructed to assign one of the following predefined labels, derived from the architectural styles most frequently associated with DDD implementations in the literature \cite{evans_domain-driven_2004, vernon_implementing_2013, buschmann_pattern-oriented_1996}, based on the structural evidence observed in the file directory, source code, and documentation: \textit{Layered Architecture}, \textit{Clean Architecture}, \textit{Hexagonal Architecture}, \textit{Onion Architecture}, \textit{CQRS}, \textit{Event-Driven Architecture}, or \textit{Microservices}. Repositories that exhibited mixed or insufficiently distinct structural signals were labelled as \textit{Unclassified}.

Architectural labels were derived using the same triplicate majority-vote strategy applied to the binary DDD classification. Specifically, the architectural style reported for each repository corresponds to the label that appeared in at least two out of three independent classification runs. In cases where all three 
runs produced different architectural labels (a full disagreement), the repository was assigned the \textit{Unclassified} label to avoid introducing noise into the architectural distribution. In a small number of cases (44 repositories, 1.76\%), the model assigned labels outside the predefined taxonomy, such as \textit{Modular Monolith} or \textit{Vertical Slice Architecture}. These were grouped as \textit{Other/Unclassified} in the final distribution to maintain consistency with the defined label set. The architectural label columns (\texttt{Architecture1}, \texttt{Architecture2}, \texttt{Architecture3}) recorded in our database enabled post-hoc agreement analysis, confirming that the majority-vote consensus rate for architectural labels was consistent with the near-perfect stability observed for the binary DDD label ($\kappa \geq 0.96$).

\subsection{Reliability of the Classification Approach} \label{subsec:reliability}
To ensure the robustness of our dataset, we evaluated the classification approach along two dimensions: the internal consistency of the LLM across repeated runs (stability), and its external validity against human judgement (accuracy).

\subsubsection{Internal Consistency (Stability)}
Given the speculative nature of LLMs, it is critical to verify that the classification results are reproducible and not artifacts of randomness. We analysed the agreement metrics across the three independent annotation passes performed for each repository.

The model demonstrated near-perfect stability. The \textit{Unanimous Agreement Rate} (where all three runs produced the exact same label) was \textbf{94.75\%}. The \textit{Majority Agreement Rate} (where at least two out of three runs agreed) was \textbf{99.95\%}, with a Disagreement Rate (a 2–1 split) of only 5.20\%. There were no instances where the model failed to reach a majority consensus.

To quantify this stability beyond simple percentage agreement, we calculated Cohen’s Kappa ($\kappa$) pairwise between the runs. The results indicated high labeling stability:
\begin{itemize}
 \item Run 1 vs. Run 2: $\kappa = 0.96$
 \item Run 1 vs. Run 3: $\kappa = 0.97$
 \item Run 2 vs. Run 3: $\kappa = 0.97$
\end{itemize}
Across all passes, pairwise Cohen’s $\kappa$ for the binary DDD label consistently ranged from \textbf{0.96 to 0.97}. These values indicate near-perfect agreement, confirming that the \textit{Triplicate Validation Strategy} successfully mitigated the non-deterministic variance of the model.

It is important to note that architectural classification from source code alone is inherently more challenging than binary DDD validation; boundaries between styles such as Layered and Clean Architecture are not always structurally unambiguous. Accordingly, the unanimous agreement rate for architectural labels 
(88.05\%) was somewhat lower than the 94.75\% observed for the binary DDD label, which is consistent with the increased subjectivity inherent in multi-class structural classification. We therefore treat the resulting distribution as indicative of dominant structural tendencies within the ecosystem rather than definitive architectural ground truth.

\subsubsection{External Validity (Human vs. LLM)}
To verify the accuracy of these stable predictions, we conducted a validation study against a human-established ground truth, following a methodology consistent with prior MSR landscape studies \cite{nikoo_empirical_2025}. We selected a random sample of 50 repositories from the candidate pool of 4,206. Two authors independently inspected the source code, folder structures, README contents, and GitHub topic labels of each repository to manually determine whether it represented a genuine DDD implementation, applying the same liberal threshold used in the LLM system prompt: a repository was labelled \textit{YES} if clear structural DDD intent was observable, even if the implementation was not theoretically complete.

The inter-rater agreement between the two human assessors was calculated using Cohen's Kappa ($\kappa$) \cite{kvalseth_note_1989}. The two raters agreed on 45 out of 50 repositories (90.0\%), yielding $\kappa = 0.77$, which indicates \textbf{substantial} agreement \cite{landis_measurement_1977}. The 5 disagreements were concentrated on boundary cases, namely repositories that referenced DDD concepts in their documentation but exhibited only partial structural implementation. These were resolved through joint discussion and source code inspection to establish a verified set of 50 labelled repositories. The YES rates of the two raters (68\% and 70\%) remained within 2 percentage points of each other, confirming calibration consistency across assessors.

We then benchmarked the majority-vote consensus label produced by the LLM pipeline against this set, treating the primary assessor's labels as the reference. The pipeline agreed with the primary assessor on 45 out of 50 repositories, achieving an agreement rate of 90.0\% ($\kappa = 0.77$, substantial \cite{landis_measurement_1977}). The model demonstrated a precision of 93.9\%, a recall of 91.2\%, and an F1-score of 92.5\%. The LLM's YES rate (66\%) was within 2 percentage points of both human raters (68\% and 70\%), indicating no systematic over- or under-classification bias. In 5 cases the LLM disagreed with both human raters simultaneously: it classified 3 repositories as NO where both humans said YES, and 2 repositories as YES where both humans said NO, showing no directional bias. These results confirm that the LLM pipeline serves as a reliable proxy for manual expert inspection, justifying its application to the full candidate dataset of 4,206 repositories.

\begin{table}[htbp]
\centering
\small
\resizebox{\columnwidth}{!}{%
\begin{tabular}{lrrrrrrr}
\hline
\textbf{Comparison} & \textbf{Agree.} & $\boldsymbol{\kappa}$ & \textbf{TP} & \textbf{TN} & \textbf{FP} & \textbf{FN} & \textbf{F1} \\
\hline
A1 vs.\ A2 (inter-rater) & 45/50 (90\%) & 0.77 & 32 & 13 & 2 & 3 & 92.8\% \\
LLM vs.\ A1 & 45/50 (90\%) & 0.77 & 31 & 14 & 2 & 3 & 92.5\% \\
LLM vs.\ A2 & 40/50 (80\%) & 0.54 & 29 & 11 & 4 & 6 & 85.3\% \\
\hline
\multicolumn{8}{l}{\small A1 = Assessor 1 (YES: 68\%), A2 = Assessor 2 (YES: 70\%), LLM (YES: 66\%)} \\
\hline
\end{tabular}%
}
\caption{Inter-rater agreement and LLM validation on the set ($n = 50$).
A1 = Assessor 1, A2 = Assessor 2. $\kappa$ is substantial (${\geq}0.61$) for A1 vs.\ A2
and LLM vs.\ A1, and moderate (${\geq}0.41$) for LLM vs.\ A2, per Landis and Koch
\cite{landis_measurement_1977}. TP/TN/FP/FN use A1 as reference for LLM rows.
F1 is for the YES class.}
\label{tab:kappa_results}
\end{table}

\subsection{Data Extraction and Synthesis} \label{subsec:data_extraction_and_synthesis}
To ensure a rigorous and structured analysis of the verified DDD landscape, we implemented a comprehensive data extraction and synthesis workflow. Our process involved the retrieval of high-fidelity metadata, technical logs, and source artifacts, which were subsequently processed through both quantitative and qualitative lenses.

\subsubsection{Data Extraction Process}
The data extraction was performed using a custom-developed \textit{Python} tool designed to interact with the collected data. For each of the \textbf{2,502 verified repositories}, we extracted a multi-dimensional dataset that captures the technical, social, and temporal facets of the projects. The extracted data was organized into a normalized \textit{SQLite} database, which served as the primary artifact for our subsequent analysis. The main categories of extracted data included:

\begin{itemize}
 \item \textbf{Repository Metadata:} We collected unique identifiers, ownership details (Individual vs. Organization), and fundamental indicators of project scale, such as the total number of stars, forks, and watchers.
 \item \textbf{Development Dynamics (Commits and Code):} For every project, we retrieved complete commit logs, including author information, timestamps, and commit messages. We also extracted file-level statistics using the \textit{loc\_data} entity to quantify lines of code (LOC), number of files, and the distribution of programming languages.
 \item \textbf{Collaboration Artifacts (Issues and PRs):} We extracted detailed records for issues and pull requests, including their creation and closure/merge dates, author information, and the textual content of associated comments.
 \item \textbf{Project Documentation:} The contents of README files and user-defined topic labels were extracted to support business domain classification and the identification of tactical design patterns.
\end{itemize}

Complementing the API-based extraction, we cloned the default branch of each verified repository. This local snapshot allowed us to verify metadata consistency and perform an in-depth analysis of the project's folder structures and implementation artifacts.

\subsubsection{Data Synthesis and Statistical Methods}
To answer our research questions ($RQ_1$ to $RQ_6$), we applied a mixed-methods synthesis strategy. We differentiated our approach based on the nature of the data:

\textbf{Quantitative Synthesis.} For $RQ_1$ (Temporal Evolution), $RQ_2$ (Architecture), $RQ_4$ (Ownership), $RQ_5$ (Ecosystems), and $RQ_6$ (Sustainability), we utilized descriptive statistics, including frequencies, percentages, and mean/median values. Temporal trends were synthesized by aggregating commit activity and project creation dates over a 21-year window to identify industrial inflection points. Business domains in $RQ_5$ were synthesized by mapping projects to the 16-domain business taxonomy established by Saeedi Nikoo et al. \cite{nikoo_empirical_2025}, whose categories are aligned with the GICS \cite{msci_gics_2024} and the TRBC \cite{refinitiv_trbc_2020}. Project sustainability in $RQ_6$ was synthesized through metrics such as PR merge latency (calculated as the duration between PR creation and merge) \cite{kochanthara_painting_2022, cosentino_systematic_2017} and repository longevity (the duration between project creation and the most recent commit) \cite{kalliamvakou_promises_2016, liao_prediction_2019}.

\textbf{Qualitative Synthesis.} To address $RQ_3$ (Exemplary Projects), we performed a qualitative synthesis following established empirical software engineering research guidelines \cite{ralph_empirical_2020}. We focused on the top 10 most technically intensive projects, identified by total commit volume as a proxy for technical maturity and architectural evolution, among repositories filtered to have more than two distinct contributors to exclude personal repositories and bots, thereby ensuring that only genuinely collaborative, engineered systems were considered \cite{kalliamvakou_promises_2016, munaiah_curating_2017}. Our synthesis involved an independent manual inspection of these repositories' source code and documentation to characterize their tactical fidelity, the use of Bounded Contexts, and their specific architectural implementations. This deep-dive allowed us to extract representative ``engineered'' characteristics that distinguish industrial DDD adoption from personal experiments.

\section{Results} \label{sec:results}
In this section, we present the empirical findings of our landscape study, derived from the systematic analysis of over 2,500 repositories verified through our automated architectural assessment pipeline. By integrating repository metadata, commit-level logs, and LLM-verified architectural labels, we provide a quantitative characterisation of how DDD is implemented and sustained within the open-source community.

Before presenting the results of our six research questions, we report the empirical noise rate analysis of our hybrid discovery strategy, as this contextualises the scale of label noise that the semantic validation pipeline was designed to address.

Table~\ref{tab:discovery_noise} presents the breakdown of candidates, verified repositories, and noise rates per discovery query.

\begin{table}[htbp]
\centering
\begin{tabular}{lrrr}
\hline
\textbf{Discovery Query} & \textbf{Candidates} & \textbf{Verified} & \textbf{Noise Rate} \\ \hline
README: \texttt{domain-driven-design} & 6,831 & 1,181 & 82.7\% \\
Topic: \texttt{ddd} & 1,995 & 720 & 63.9\% \\
Topic: \texttt{domain-driven-design} & 1,443 & 521 & 63.9\% \\
README: \texttt{ddd} & 1,350 & 69 & 94.9\% \\
Topic: \texttt{value-object} & 89 & 5 & 94.4\% \\
Other specific topics & 34 & 6 & 82.4\% \\
\hline
\textbf{Total} & \textbf{11,742} & \textbf{2,502} & \textbf{78.7\%} \\
\hline
\end{tabular}
\caption{Noise rate per discovery query source. Noise rate is defined as the percentage of candidates that did not survive semantic validation. Low-frequency specific topics (\texttt{domain-event}, \texttt{domain-driven-development}, \texttt{aggregate-root}) are aggregated in the ``Other specific topics'' row for brevity; their individual counts are: \texttt{domain-event} (8 candidates, 3 verified), \texttt{domain-driven-development} (13 candidates, 2 verified), \texttt{aggregate-root} (3 candidates, 1 verified).}
\label{tab:discovery_noise}
\end{table}

These results show that keyword-based discovery alone is dominated by false positives, with an overall noise rate of 78.7\%. Crucially, the noise is highly uneven across sources: author-assigned \textit{topic} labels (around 64\% noise) are considerably more reliable than free-text \textit{README} matches, and the short abbreviation query \texttt{in:readme ddd} is almost entirely noise (94.9\%, fewer than one in twenty candidates genuine). The takeaway is that, for an ambiguous domain abbreviation such as ``DDD'', metadata-based retrieval cannot on its own yield a trustworthy dataset, which is precisely the gap the semantic validation pipeline is designed to close.

\subsection{(RQ\textsubscript{1}): How has the prevalence and activity of DDD-related repositories on GitHub evolved since its earliest appearance in open source, and is the interest sustained?} \label{subsec:rq1_results}
To understand the genesis and growth of DDD in the open-source ecosystem, we analysed the creation dates of verified repositories and their aggregate commit activity over a 21-year window. Following the methodology of previous landscape studies \cite{kochanthara_painting_2022, rigas_mining_2023}, this measurement serves as a barometer of the vibrancy of the community and the progress of the paradigm. Furthermore, we incorporated repository longevity defined as the duration between a project's creation and its most recent commit to distinguish actively developed, engineered systems from temporary experiments or ``code dumps'' often abandoned shortly after creation \cite{kalliamvakou_promises_2016}.

\subsubsection{Project Creation Trends}
Our data indicates that the entry of verified DDD projects into GitHub remained sparse for nearly a decade following the paradigm's introduction by Evans in 2004 \cite{evans_domain-driven_2004}. The genesis of the currently active landscape began in 2009 with a single repository, followed by a period of infancy where annual project creation remained in the single or double digits until 2016. 

\begin{figure}[htbp]
 \centering
 \includegraphics[width=\textwidth]{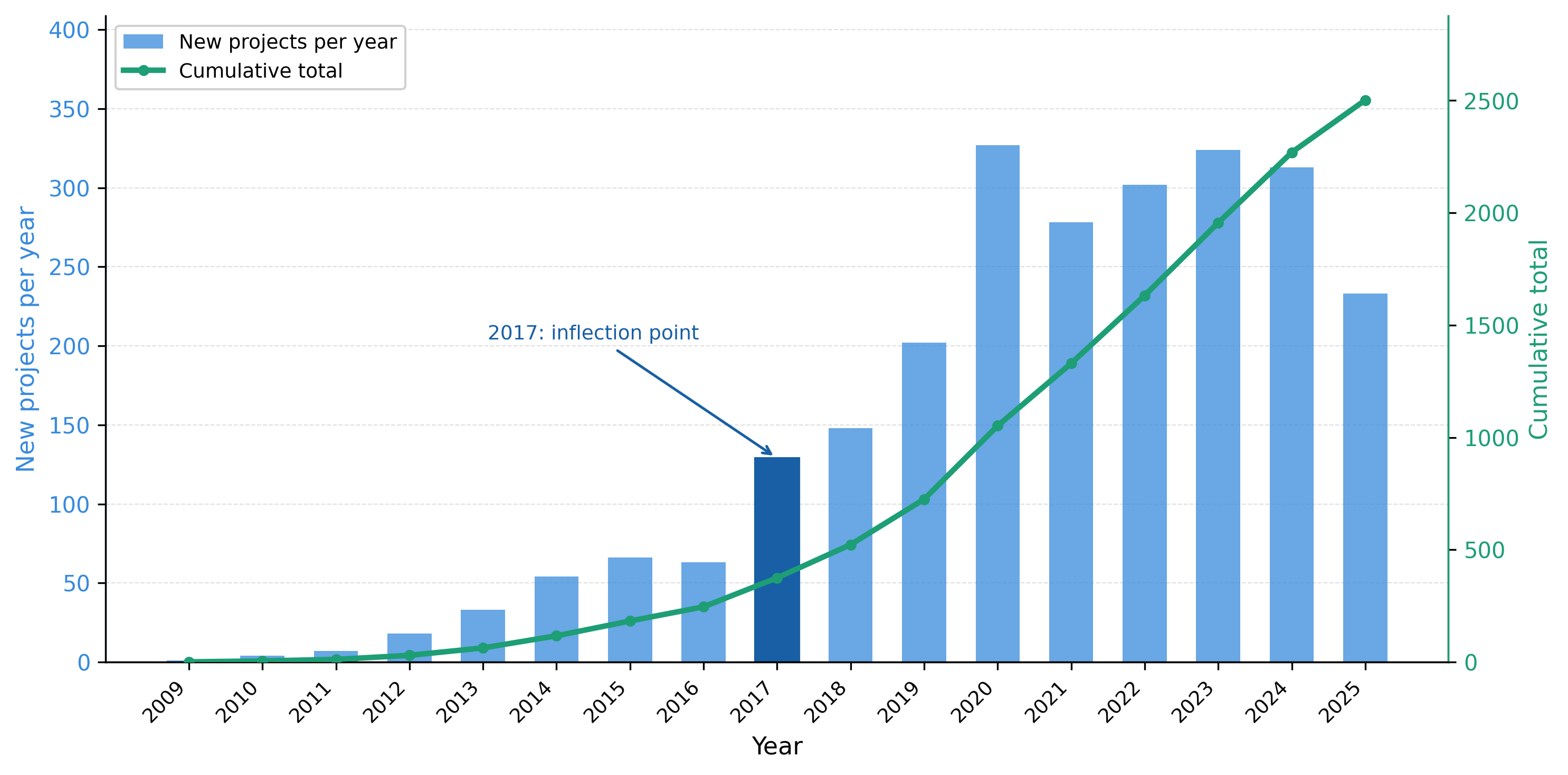}
 \caption{Annual creation of verified DDD repositories on GitHub (bars, left axis) and cumulative total (line, right axis), from 2009 to 2025. The 2017 inflection point marks the transition from sparse infancy to sustained open-source adoption. The 2025 figure reflects a partial year of data collection.}
 \label{fig:project_creation_trends}
\end{figure}

As shown in Figure~\ref{fig:project_creation_trends}, a significant turning point occurred in 2017, when the number of new verified projects surged to 129, a trend that has since exhibited sustained acceleration. Between 2019 and 2020, the creation rate increased by over 60\% (from 202 cases to 327), mirroring the boom patterns observed in other emerging technical domains like automotive software \cite{kochanthara_painting_2022, rigas_mining_2023}. Creation reached a peak in 2020 with 327 new projects and from 2020 to 2024, the community has consistently added over 300 verified DDD projects annually, suggesting that interest in implementing rich domain models is both robust and sustained. While the underlying causes cannot be established from repository metadata alone, this pattern is consistent with the broader increase in open-source development activity observed globally during that period \cite{github_octoverse_2020}.

\subsubsection{Development Activity and Maturity}
The aggregate commit activity provides a deeper view of development dynamics beyond simple repository creation. Interestingly, our logs show commit activity as early as 2004 (169 commits), which likely reflects the peril of imported version history from older systems or mirrors predating the official creation of GitHub accounts \cite{kalliamvakou_promises_2016}. 

The technical intensity of the field reached a critical mass in 2017, coinciding with the rise in project creation. Aggregate commits tripled between 2016 and 2018 (increasing from 18,109 to 34,996) and reached a historical peak in 2023 with 98,151 commits. This sustained activity suggests that DDD projects are not merely code dumps or temporary experiments, which typically exhibit low activity or are abandoned within 24 hours \cite{kalliamvakou_promises_2016}. Instead, the long-term trend of increasing insertions and file modifications indicates that verified DDD repositories represent actively maintained, engineered systems with substantial software content \cite{kalliamvakou_promises_2016, rigas_mining_2023}

\subsubsection{Repository Longevity and Continuity}
Repository longevity is a vital indicator of project continuity and community interest \cite{kalliamvakou_promises_2016, liao_prediction_2019}. The verified DDD projects exhibit an average longevity of 660.73 days and a median longevity of 340.37 days, with the most enduring project reaching a maximum longevity of 5,438.57 days (approx. 14.9 years). The minimum longevity was recorded as -1,697.07 days, a known artefact of imported version histories in GitHub repositories \cite{kalliamvakou_promises_2016}. The median longevity of 340.37 days is substantially higher than the 9.9-day median reported for the broader GitHub population \cite{kalliamvakou_promises_2016}.

\subsection{(RQ\textsubscript{2}): What is the distribution of high-level architectural styles within the verified DDD ecosystem?}

To provide a technical characterisation of the DDD landscape, we utilised our majority-voting classification pipeline (Section \ref{subsec:semantic_classification}) to categorise the verified repositories at scale based on their dominant structural patterns.

\subsubsection{Dominant Architectural Styles}
The empirical distribution of architectural styles, derived from our majority-vote classification pipeline (Section~\ref{subsec:semantic_classification}), indicates that Layered Architecture is the most prevalent pattern, appearing in 28.9\% (723 repositories) of the verified DDD ecosystem, followed by Clean Architecture at 22.78\% (570 repositories). Table~\ref{tab:architectural_distribution} presents the full distribution across all identified styles.

\begin{table}[htbp]
\centering
\begin{tabular}{lrr}
\hline
\textbf{Architectural Style} & \textbf{Count} & \textbf{Percentage} \\ \hline
Layered Architecture & 723 & 28.90\% \\
Clean Architecture & 570 & 22.78\% \\
Hexagonal Architecture & 298 & 11.91\% \\
CQRS & 285 & 11.39\% \\
Onion Architecture & 276 & 11.03\% \\
Event-Driven Architecture & 242 & 9.67\% \\
Microservices & 64 & 2.56\% \\
Other / Unclassified & 44 & 1.76\% \\ \hline
\textbf{Total} & \textbf{2,502} & \textbf{100\%} \\ \hline
\end{tabular}
\caption{Distribution of architectural styles across the 2,502 verified DDD repositories ($RQ_2$), derived from the majority-vote GPT-4o classification pipeline.}
\label{tab:architectural_distribution}
\end{table}

\subsubsection{Emergent Patterns}
A secondary group of architectural styles shows a relatively uniform distribution, consisting of Hexagonal Architecture (11.91\%), CQRS (11.39\%), and Onion Architecture (11.03\%). More operationally complex styles, such as Event-Driven Architecture (9.67\%) and Microservices (2.56\%), constitute a smaller segment of the landscape. Taken together, Layered and Clean architectures account for over half of the verified ecosystem, indicating that open-source DDD is predominantly implemented through conventional, well-documented structural templates. The more specialised distributed styles often associated with DDD in the literature, such as Event-Driven Architecture and Microservices, remain a minority in practice, pointing to a gap between the architectural patterns emphasised in research and those most commonly adopted in open source.

\subsection{(RQ\textsubscript{3}): What are the characteristics of the most technically intensive DDD projects identified by the volume of sustained commit activity, and what business domains and architectural patterns do they showcase?} \label{subsec:rq3_results}

Applying the selection procedure described in Section~\ref{subsec:data_extraction_and_synthesis}, we manually examined the top 10 most \textit{technically intensive} DDD projects in depth. Three consistent themes emerged across this set, which we present in turn: their structural characteristics and tactical fidelity, their dominant architectural patterns, and the business domains they represent.

\subsubsection{Characteristics and Tactical Fidelity}
The top 10 technically intensive projects exhibit a high degree of structural modularity and tactical fidelity. A recurring characteristic is the explicit division of the system into multiple Bounded Contexts (e.g., \texttt{shortlink-org/shortlink}, \texttt{smartstore/Smartstore}), facilitating parallel development by allowing teams to work independently on self-contained modules \cite{ozkan_domain-driven_2025, nikoo_empirical_2025}. Our findings show that these projects consistently showcase Rich Domain Models rather than anemic ones, mirroring the core principles of the paradigm \cite{evans_domain-driven_2004}.

For instance, projects like \texttt{generike/SmartStoreNET} and \texttt{kurrent-io/KurrentDB} encapsulate complex business rules, such as inventory management invariants and cluster node state transitions, directly within Entities and Aggregates. The use of Value Objects to enforce domain-specific constraints (e.g., \textit{vo\_url} for link validation) is pervasive, supporting the observation that Entities and Value Objects remain the most extensively adopted tactical patterns in industrial settings \cite{vernon_implementing_2013, hippchen_designing_2017}. Furthermore, the presence of comprehensive documentation in repositories like \texttt{AxonFramework/AxonFramework} and \texttt{RailsEventStore/rails\_event\_store} serves as a vital tool to mitigate the high learning curve typically associated with the paradigm \cite{ozkan_domain-driven_2025}.

\subsubsection{Dominant Architectural Patterns}
Our analysis reveals that the most active DDD projects often move beyond simple Layered Architecture toward more sophisticated styles designed for distributed environments. CQRS (Command Query Responsibility Segregation) and Event Sourcing emerged as dominant architectural patterns among the top tier (e.g., \texttt{AxonFramework/AxonFramework}, \texttt{shortlink-org/shortlink}). This aligns with the observation that interest in DDD has peaked since 2017, specifically due to its prominence in facilitating Microservices decomposition and managing consistency in distributed data-intensive systems \cite{ozkan_domain-driven_2025, krause_microservice_2020, maddodi_aggregate_2020}.

Additionally, several top projects function as frameworks or boilerplate systems (e.g., \texttt{apache/causeway}, \texttt{aspnetboilerplate/aspnetboilerplate}). These projects provide the underlying infrastructure, such as base classes for \textit{AggregateRoot} and \textit{DomainService}, to reduce the cognitive load of implementation. This trend reflects the increasing focus on ``tooling support'' to close the model-code gap in complex software projects \cite{ozkan_domain-driven_2025, kapferer_domain-driven_2020}.

\subsubsection{Business Domain Showcase}
When mapped to the 16-domain industry taxonomy \cite{nikoo_empirical_2025}, the most technically intensive projects are primarily concentrated in the \textit{Traditional Software} and \textit{Sales} sectors. This confirms that DDD serves as a stable tool for sectors characterized by high transactional complexity and the need for rigorous domain integrity to maintain competitive advantage \cite{evans_domain-driven_2004, ozkan_domain-driven_2025}. Projects like \texttt{smartstore/Smartstore} demonstrate how DDD manages the ``Core Domain'' of e-commerce, handling multifaceted business rules for catalogs, payments, and shipping. The prevalence of software frameworks, in turn, reflects the paradigm's role in building robust developer tools that serve as the ``heart'' of other software-intensive industries \cite{evans_domain-driven_2004, ozkan_domain-driven_2025}. In Section \ref{subsubsec:business_landscape_domain_identification}, we took a closer look at the individual business domains.

\subsection{(RQ\textsubscript{4}): What is the distribution of repository ownership and how does it influence the rigour of DDD adoption?} \label{subsec:rq4_results}
To characterise the organisational structure of the DDD ecosystem, we analysed the distribution of repository ownership between Individual Users and Organisations. Ownership type indicates whether a paradigm is driven primarily by individual developers or by collective, team-based efforts. We note that organisation ownership on GitHub spans companies, academic groups, and open-source foundations, and therefore does not by itself denote industrial adoption \cite{kochanthara_painting_2022, cosentino_systematic_2017}.

\subsubsection{Ownership Distribution}
The empirical results from our verified dataset show that 76.98\% (1,926 repositories) are owned by individual users, while 23.02\% (576 repositories) are owned by organisations. This distribution suggests that while DDD has a significant personal and educational following, nearly one in four verified projects is maintained under an organisational umbrella. This organisational footprint (approx. 23\%) is notably higher than that observed in other specialised modelling domains such as BPMN (16\%) \cite{nikoo_empirical_2025}, yet it remains below the one in three benchmark observed in high-integrity engineering fields like automotive software (33\%) \cite{kochanthara_painting_2022, cosentino_systematic_2017}.

\subsubsection{Influence on Implementation Rigour}
The literature on the perils of GitHub data suggests that individual repositories are often susceptible to being code dumps, personal experiments, or non-engineered systems \cite{kalliamvakou_promises_2016}. However, because our dataset consists exclusively of projects that passed the LLM-based semantic verification, even the 76.98\% of user-owned repositories demonstrate a clear intent to implement tactical and strategic DDD patterns, such as Aggregates, Value Objects, and Bounded Contexts.

As identified in prior landscape studies, organisation-owned projects typically solicit higher levels of internal and external participation and exhibit more sustained development activity compared to user projects \cite{nikoo_empirical_2025}. The presence of 576 organisationally-backed DDD projects within the verified open-source ecosystem suggests that the paradigm extends beyond individual experimentation into structured team-based development, providing a basis for further investigation into its adoption in organisational and team-based contexts.

\subsection{(RQ\textsubscript{5}): What are the dominant technological and business ecosystems within the verified DDD landscape?} \label{subsec:rq5_results}

DDD is intended to manage complexity in the heart of software by aligning technical artifacts with business requirements \cite{evans_domain-driven_2004}. To map the practical application of this paradigm, we analysed the technological preferences and business sectors of the repositories. This characterisation serves as a picture of which ecosystems are currently leading the practical adoption of DDD principles \cite{rigas_mining_2023}.

\subsubsection{Technological Landscape: Programming Language Distribution}
Our analysis reveals a diverse but highly concentrated technological ecosystem. C\# is the dominant programming language, used in 34.17\% (855 repositories) of the verified dataset. This prevalence aligns with the historically strong influence of the .NET community in advocating for DDD, as seen in foundational practitioner literature \cite{vernon_implementing_2013}. Following C\#, TypeScript represents the second most prevalent language with 443 repositories (17.71\%), closely followed by Java with 390 repositories (15.59\%).

Interestingly, while the DDD academic literature has historically focused on Java-based examples \cite{ozkan_domain-driven_2025}, our data indicates a more diverse industrial landscape. Modern ecosystems like Go (199 repositories) and Python (134 repositories) have established a stable presence, though they remain secondary to established enterprise-grade languages like C\# and Java. This profile diverges from general programming-language popularity indices: as of early 2026, Python leads the TIOBE Index whereas C\# ranks around fifth \cite{tiobe_index_2026}, yet within our verified DDD corpus, C\# is the dominant language and Python only secondary. This inversion suggests that DDD adoption concentrates in statically typed, enterprise-oriented ecosystems rather than tracking overall language popularity. In contrast to specialized domains like automotive software, which have seen a total shift from MATLAB to Python \cite{kochanthara_painting_2022, cosentino_systematic_2017}, the DDD community remains anchored in languages that traditionally support rich type systems and complex object-oriented patterns required for implementing Aggregates and Value Objects \cite{evans_domain-driven_2004, ozkan_domain-driven_2025}.

\subsubsection{Business Landscape: Domain Identification} \label{subsubsec:business_landscape_domain_identification}
Using the rule-based keyword classification described in Section~\ref{subsec:data_extraction_and_synthesis}, each of the 2,502 verified repositories was mapped to one of 16 business domains. The verified DDD ecosystem is heavily concentrated in Traditional Software, which accounts for 43.96\% (1,100 repositories) of the dataset, followed by 25.30\% (633 repositories) that could not be assigned an explicit business domain. This reflects the paradigm's inherent focus on managing technical complexity within software-intensive systems \cite{evans_domain-driven_2004}.

Despite the improved multi-field search (incorporating names, descriptions, and README files), 25.3\% (633 repositories) remained classified as \textit{Unknown/Other}, reflecting the limits of keyword-based domain classification against sparse repository metadata \cite{kalliamvakou_promises_2016, munaiah_curating_2017}.

Among identifiable commercial sectors, Media \& Publishing (8.35\%) and Financial Services (6.12\%) emerged as the sectors with the highest adoption rates. This confirms that DDD is a stable tool for industries characterised by high transactional complexity and the need for rigorous domain integrity \cite{evans_domain-driven_2004, guemes_emerging_2018}. Conversely, sectors like Agriculture (0.12\%) and Government Services (0.48\%) remain underrepresented. 

\subsection{(RQ\textsubscript{6}): How does the community interact with DDD projects in terms of popularity and maintenance dynamics, such as PR latency and activity frequency?} \label{subsec:rq6_results}

To evaluate the vibrancy and sustainability of the verified DDD landscape, we analysed community reception through popularity metrics and maintenance responsiveness through PR merge latency. 

\subsubsection{Community Reception and Popularity}
Our empirical analysis reveals a wide variation in community reception across the verified DDD ecosystem. The projects in our dataset have an average of 132.07 stars and 25.88 forks, though the median values of 5 stars and 1 fork more accurately represent the typical repository, reflecting a right-skewed distribution where a small number of highly prominent projects pull the averages upward. While these figures are in the hundreds and tens rather than thousands, they represent a significant level of engagement when compared to the long tail of GitHub, where the majority of repositories are inactive, personal code dumps with near-zero social interaction \cite{kalliamvakou_promises_2016}.

For context, these averages are markedly higher than the baseline observed in emerging technical landscapes such as automotive software, where the median number of stars and forks was recorded at 24 and 9, respectively \cite{kochanthara_painting_2022, cosentino_systematic_2017}. Furthermore, the verified DDD repositories exhibit a high level of collaborative participation, with an average of 11.15 contributors per project. As noted in the literature, projects with more than two committers are more likely to be actively-developed, engineered systems rather than temporary experiments or academic toys. This level of reception confirms that the verified DDD landscape serves as a picture of a vibrant and progress-oriented community \cite{rigas_mining_2023}.

\subsubsection{Maintenance Practices and Merge Latency}
The sustainability of the DDD ecosystem is further evidenced by the responsiveness of project maintainers to external contributions. We measured this via PR merge latency, which reflects the technical intensity and maintenance health of the repositories \cite{kochanthara_painting_2022, cosentino_systematic_2017}. Our data shows that the verified 
DDD ecosystem maintains an average merge latency of 3.62 days and a median merge latency of 0.03 days (approximately 49 minutes). The substantial gap between the mean and median indicates a right-skewed distribution: the majority of pull requests are merged within the hour, while a small number of large or 
controversial contributions with extended review cycles pull the average upward.

The median latency of under one hour is the more representative measure of day-to-day maintenance responsiveness in this ecosystem. It suggests that project maintainers operate with a high degree of trust in their contributor base, merging straightforward contributions rapidly while reserving extended 
review for architecturally significant changes. This pattern is consistent with the collaborative maturity observed in the contributor counts ($RQ_6$) and further distinguishes the verified DDD landscape from the typical GitHub peril of sparse activity or project abandonment \cite{kalliamvakou_promises_2016}.

\section{Discussion} \label{sec:discussions}

In this section, we synthesize our empirical findings from $RQ_{1}$ through $RQ_{6}$ to provide a comprehensive answer to our main research question ($RQ_{M}$): \textit{What are the characteristics and trends of DDD adoption in open-source projects on GitHub?} We frame this discussion through three thematic lenses: (1) the inflection point of industrial maturity, (2) architectural alignment with distributed systems, and (3) a methodological reflection on the documentation ``peril'' in engineered software.

\subsection{The 2017 Inflection Point and Industrial Maturity}
Our trend analysis ($RQ_{1}$) identifies 2017 as a critical turning point for the DDD paradigm in open source. While the paradigm remained in a sparse ``infancy'' for over a decade following Evans' seminal work \cite{evans_domain-driven_2004}, the surge in 2017 where new verified projects doubled and aggregate commits tripled, marks the transition of DDD from a theoretical design proposal to a practical paradigm. This trend aligns with our systematic literature review, which noted a peak in research interest starting in 2017 as the community shifted focus toward modern architectural challenges like microservices \cite{ozkan_domain-driven_2025}.

Beyond mere volume, the character of the ecosystem points to professional-grade maturity. Verified projects persist far longer than the typical GitHub repository ($RQ_1$), a contrast that recent survival analyses confirm still holds \cite{ait_survival_2022}. Together with the semantic verification that excludes non-engineered candidates, this indicates a landscape of stable, engineered systems rather than short-lived experiments or code dumps \cite{kochanthara_painting_2022, munaiah_curating_2017}. The organisational share (23.02\%) is also higher than in comparable modelling domains such as BPMN (16\%) \cite{nikoo_empirical_2025}; because organisation ownership on GitHub spans companies, academic groups, and foundations, we read this as evidence of collective, team-based adoption rather than industrial practice specifically.

\subsection{Architectural Alignment: From Monoliths to Distributed Complexity}
The architectural distribution ($RQ_{2}$) and our deep-dive into technically intensive projects ($RQ_{3}$) reveal a clear trend: the modern DDD characteristic is its synergy with distributed and decoupled styles. While Layered Architecture (28.9\%) remains the most prevalent, likely due to its historical association with DDD. Clean Architecture (22.78\%) has emerged as the preferred choice for high-fidelity implementations. 

Our qualitative analysis of the most active projects shows that developers are increasingly utilizing Strategic Design patterns, such as Bounded Contexts, to facilitate parallel development in multi-stakeholder environments \cite{vernon_implementing_2013, brandolini_eventstorming_2013}. The dominance of CQRS and Event Sourcing among top-tier projects (e.g., \texttt{AxonFramework/AxonFramework}, \texttt{shortlink-org/shortlink}) highlights a specific trend where DDD is leveraged not just for logic encapsulation, but as a structural tool to manage consistency and complexity in distributed data-intensive systems \cite{ozkan_domain-driven_2025, joselyne_partitioning_2018}. This mirrors the literature's finding that microservices received the highest research attention (44\%) due to DDD’s ability to provide a rigorous "heart" for service decomposition \cite{ozkan_domain-driven_2025}.

\subsection{Industrial Adoption: Technological Preferences and the Business ``Peril''}
The technological landscape of DDD ($RQ_{5}$) is characterized by a strong concentration in \textit{C\# (34.17\%)} and \textit{TypeScript (17.71\%)}. The C\# dominance reflects the historically strong advocacy for the paradigm within the .NET community \cite{vernon_implementing_2013}, while the rise of TypeScript suggests a new trend of applying DDD to modern full-stack and cloud-native environments.

Identifying the business sectors these projects serve proved difficult, however, because repository documentation tends to describe how a system is built rather than the business problem it solves \cite{almarzouq_mining_2020}. This is a substantive obstacle to domain-level analysis of engineered software, and it is the primary reason a large share of repositories resisted sectoral classification ($RQ_5$); we treat it as a threat to validity in Section~\ref{sec:limitations_threats_to_validity}. 

Where sectors were identifiable ($RQ_5$), adoption concentrated in transaction-intensive domains such as Media \& Publishing and Financial Services, consistent with DDD's emphasis on rigorous domain integrity where business rules are complex and consequential \cite{evans_domain-driven_2004, ozkan_domain-driven_2025}.

\subsection{Meta-Methodological Reflection: The Mining-Architecture Gap}
Beyond the empirical trends, our study highlights a significant meta-methodological gap within the software engineering community. As noted in the recent reflection by Robles et al. \cite{robles_reflection_2023}, there is a persistent ``culture gap'' between repository miners (MSR) and software architects (Modeling), with a minimal 1.7\% overlap in research participation across these two domains \cite{robles_reflection_2023}. 

Our findings in $RQ_{2}$ and $RQ_{3}$ confirm that while DDD provides a rigorous framework for aligning code with business logic, the \textit{intent} behind these designs is often poorly captured in Git-based versioning systems. As Robles et al. argue, these systems remain fundamentally optimized for textual code rather than conceptual modeling or architectural goals \cite{robles_reflection_2023}. This ``partial view'' provided by repository mining explains the metadata sparsity we observed in our business domain classification ($RQ_5$). To move the field beyond opinion-based argument regarding architectural efficacy and toward a truly data-driven paradigm, we join the call for better traceability standards, analogous to bug-fix linking, where architectural pattern implementations and bounded context decisions are explicitly tagged within the commit history \cite{ozkan_domain-driven_2025,robles_reflection_2023}.

This persistent gap necessitates a shift in how researchers approach architectural artifacts, leading to several implications for the future of repository mining.

\subsection{Sustainability and Community Response}
Finally, the sustainability of the ecosystem ($RQ_{6}$) is characterized by a high degree of collaborative vibrancy. With an average of \textit{11.15 contributors per project} and an average merge latency of 3.62 days (median: 49 minutes), the verified DDD landscape outperforms other specialized technical domains \cite{kochanthara_painting_2022, cosentino_systematic_2017}. This ``fast turnaround'' in pull requests suggests that project maintainers are actively engaged in code reviews and community growth, mitigating the high learning curve typically associated with the paradigm \cite{ozkan_domain-driven_2025, kalliamvakou_promises_2016}. The presence of numerous frameworks and boilerplate systems in the top-tier repositories (e.g., \texttt{aspnetboilerplate/aspnetboilerplate}, \texttt{apache/causeway}) further points to an industrial trend of creating tool support to close the ``model-code gap'' and lower the barrier for DDD adoption in practice \cite{ozkan_domain-driven_2025}.

\section{Implications} \label{sec:implications}

The results of our large-scale characterization of the DDD landscape on GitHub offer significant implications for both the research community and software engineering practitioners. We ground these implications in our empirical findings specifically the 2017 turning point, the 660-day project longevity, and the rise of modern cloud-native implementations while reflecting on the meta-methodological gaps identified in the literature.

\subsection{Implications for Research}

\textbf{Bridging the Mining-Architecture Gap.} A critical implication of our study is the confirmation of a persistent ``culture gap'' between the MSR and the software architecture/modeling communities. As Robles et al. noted, only 1.7\% of researchers participate in both communities, leading to a situation where miners focus on code artifacts while architects focus on conceptual intent \cite{robles_reflection_2023}. Our findings in $RQ_2$ and $RQ_3$ show that while DDD provides a rigorous framework for aligning code with business logic, this \textit{intent} is often lost in Git-based versioning systems designed for textual code, rather than conceptual modeling \cite{robles_reflection_2023}. Future research should investigate better traceability standards, analogous to linking bug IDs to commits where architectural pattern implementations and strategic Bounded Context decisions are explicitly tagged within the commit history \cite{ozkan_domain-driven_2025,robles_reflection_2023}. To reduce cognitive load, future context-mapping tools should explore automating the generation of these architectural commit tags.

\textbf{Targeting Engineered Systems in Repository Mining.} Our longitudinal analysis ($RQ_1$) identifies a median project longevity of 340.37 days, more than 34 times the 9.9-day median found in the seminal study by Kalliamvakou et al. \cite{kalliamvakou_promises_2016}. This implies that the verified DDD ecosystem is a primary source for studying ``engineered software projects'' rather than experimental ``toys'' or code dumps \cite{munaiah_curating_2017}. Researchers looking for stable, professional-grade software for longitudinal studies on software evolution, technical debt, or architectural decay should look to this ecosystem as a verified benchmark for industrial maturity \cite{kochanthara_painting_2022, nikoo_empirical_2025}.

\textbf{The Methodological Shift toward Semantic Validation.} Our use of Large Language Models (LLMs) to perform architectural verification at scale addresses a long-standing ``peril'' in GitHub mining: label noise \cite{kalliamvakou_promises_2016, tutko_how_2022}. Keyword-based discovery alone is insufficient for architectural research; semantic verification is required to distinguish true ``Rich Domain Models'' from anemic ones \cite{ozkan_domain-driven_2025}. This methodological approach serves as a blueprint for future domain specific landscape studies such as those in Electric Vehicles or BPMN to move beyond metadata and analyze actual implementation fidelity, addressing the reproducibility crisis noted by Tutko et al. \cite{tutko_how_2022, rigas_mining_2023, nikoo_empirical_2025}.

\textbf{Noise Rate Benchmarks for Domain-Specific Repository Discovery.} Our discovery source analysis provides the first empirically grounded noise rate benchmarks for domain-specific repository mining using hybrid keyword strategies. Across our full candidate pool of 11,742 repositories, the overall noise rate was 
78.7\%, with README-based queries performing substantially worse (84.72\%) than topic-based queries (64.84\%). Most critically, short abbreviation queries in free-text README searches proved highly unreliable: our \texttt{in:readme ddd} query yielded a noise rate of 94.9\%, meaning fewer than one in twenty candidates 
represented a genuine DDD implementation. We argue that these figures serve as a transferable calibration benchmark for future domain-specific landscape studies. Researchers mining repositories using short or ambiguous domain abbreviations in README searches, such as \texttt{bpm}, \texttt{iot}, or \texttt{ml}, should 
anticipate similarly high noise rates and treat semantic validation not as an optional quality step but as a methodological necessity. This finding directly extends the reproducibility concerns raised by Tutko et al.~\cite{tutko_how_2022} by providing the first concrete noise rate quantification for this class of 
retrieval query.

\subsection{Implications for Practice}

\textbf{Reference Architectures for Distributed Systems.} Our qualitative deep dive ($RQ_3$) identifies specific top tier projects, such as \texttt{AxonFramework/AxonFramework} and \texttt{generike/SmartStoreNET}, that demonstrate high fidelity implementations of CQRS and Event Sourcing \cite{ozkan_domain-driven_2025}. For practitioners, these repositories serve as ``engineered'' reference architectures for managing consistency and complexity in distributed, data-intensive environments. The prevalence of frameworks and boilerplate systems (e.g., \textit{aspnetboilerplate}, \textit{causeway}) suggests that the industry is actively moving to close the ``model-code gap'' by providing pre-built infrastructure for tactical DDD patterns \cite{ozkan_domain-driven_2025}.

\textbf{Technological Diversification Beyond Java.} Practitioners and educators should note the significant shift in technological preferences within the DDD landscape ($RQ_5$). While academic literature has historically focused on Java, our data shows that C\# (34.17\%) and TypeScript (17.71\%) are the dominant languages for industrial DDD adoption on GitHub \cite{ozkan_domain-driven_2025}. This finding challenges the ``Java-centric academic tradition'' and implies that DDD is no longer confined to the back-end but is a viable paradigm for modern full-stack and cloud-native development where strong type systems can enforce domain invariants \cite{vernon_implementing_2013}.

\textbf{Mitigating the High Learning Curve.} The sustainability metrics in $RQ_6$, specifically the high average contributor count (11.15) and fast PR merge latency (3.62 days), suggest that successful DDD projects are characterized by vibrant, responsive communities. However, the systematic literature review highlights that a \textit{high learning curve} remains a primary barrier to adoption \cite{ozkan_domain-driven_2025}. Organizations should look to the top-tier ``engineered'' projects in our study as examples of how to utilize comprehensive documentation (e.g., \textit{rails\_event\_store}) and \textit{Ubiquitous Language} to facilitate onboarding and mitigate communication gaps between engineers and domain experts \cite{ozkan_domain-driven_2025, cosentino_systematic_2017}.

\textbf{Identifying adoption leaders and competitive opportunities.} Our mapping to the 16-domain taxonomy identifies \textit{Financial Services (6.12\%)} and \textit{Sales (2.48\%)} as the primary commercial sectors for DDD ($RQ_5$). This implies that DDD has reached a level of industrial stability where it is the tool of choice for industries facing high transactional complexity \cite{evans_domain-driven_2004, guemes_emerging_2018}. Conversely, the extremely low prevalence in sectors like \textit{Agriculture (0.12\%)} and \textit{Manufacturing (2.56\%)} identifies these as under-represented sectors. Companies in these sectors may find significant competitive advantages in adopting these mature patterns to manage their own digital transformations and complex business rules \cite{ozkan_domain-driven_2025}.

\section{Limitations and Threats to Validity} \label{sec:limitations_threats_to_validity}

In this section, we address potential validity concerns following the framework by Wohlin et al. \cite{wohlin_experimentation_2012}, grounding our reflection in established MSR perils and meta-methodological gaps.

\subsection{Internal Validity}
\textbf{Metadata sparsity}. As observed in $RQ_5$, 25.3\% of repositories remained classified as \textit{Unknown/Other} because project descriptions often prioritize technical utility (e.g., ``Clean Architecture Template'') over explicit business intent. This ``partial view'' is a fundamental limitation of repository mining; Git-based systems are fundamentally optimized for tracking textual code changes rather than capturing the conceptual goals or business rationale of architectural decisions \cite{robles_reflection_2023}. While we mitigated potential classification bias through a triplicate majority-vote strategy and validated the resulting labels against an independent expert sample (Cohen's $\kappa = 0.77$, F1 = 92.5\%; Section~\ref{subsec:reliability}), the use of proprietary models like GPT-4o introduces reproducibility challenges, as model updates by the provider could theoretically alter future classification outcomes \cite{hou_large_2023}.

\subsection{Construct Validity}
\textbf{Socio-technical gap}. DDD is inherently a socio-technical paradigm involving ``Ubiquitous Language'' and ``Strategic Design'' developed through multi-stakeholder interactions such as Event Storming workshops and face-to-face meetings. Repository mining is highly effective at identifying the \textit{artifacts} of DDD (e.g., Aggregates, Bounded Contexts, or Repositories) but cannot fully capture the \textit{intent} or the collaborative process that occurred in non-versioned channels like Slack or physical workshops. This aligns with the findings of Robles et al. \cite{robles_reflection_2023}, which highlight that version control systems are not designed to store the conceptual modeling discussions that define the paradigm's success. Furthermore, by treating the repository as the unit of analysis, we may have missed activity occurring in forked ``constellations'' that represent a single logical engineered project \cite{kalliamvakou_promises_2016}.

\textbf{Domain-classification heuristic}. Our business-domain distribution ($RQ_5$) relies on keyword matching over repository names, descriptions, and README text, mapped onto the externally grounded 16-domain taxonomy of Saeedi Nikoo et al. \cite{nikoo_empirical_2025}. While the taxonomy itself is validated, the keyword-to-domain mapping is hand-crafted rather than systematically derived, and a substantial share of repositories (25.3\%) matched no domain. We therefore report the $RQ_5$ distribution as indicative of broad sectoral tendencies rather than a precise measurement; a systematic re-classification, for instance by extending our LLM pipeline to business-domain labelling, remains future work.

\subsection{External Validity}
\textbf{Selection and regional bias}. Our exclusive focus on GitHub, while justified by its status as the platform of reference, systematically excludes professional DDD implementations on GitLab, Bitbucket, or private corporate servers \cite{cosentino_systematic_2017}. Additionally, our English-language filter may underrepresent vibrant regional DDD communities (e.g., in Brazil or Europe) where documentation might not be in English, potentially biasing our sectoral results ($RQ_5$) toward globalized or North American software norms \cite{nikoo_empirical_2025}. Furthermore, our inclusion criterion of $\geq 10$ commits specifically targets \textit{engineered systems} \cite{ munaiah_curating_2017}. While this ensures high data quality, it creates a survival bias, meaning our results characterize the \textit{successful} and sustained adoption of DDD rather than the high rate of project abandonment often associated with the paradigm's steep learning curve.

\subsection{Conclusion Validity}
\textbf{Imported version history anomalies.} As identified in $RQ_1$, our logs show commit activity as early as 2004, predating the creation of many repositories. A notable statistical anomaly in our data is a minimum longevity of -1,697 days. These ``legacy commits'' result from migrating histories from older version control systems (like SVN or Mercurial) into Git, which can distort the understanding of the paradigm's temporal genesis in open source \cite{tutko_how_2022}. We have addressed this by using aggregate trends and median based analysis to prioritize the broader evolution over specific timestamp outliers.

\section{Conclusion and Future Work} \label{sec:conclusion}
In this study, we presented the first large-scale empirical characterisation of the DDD landscape within the open-source ecosystem. By mining and semantically validating 2,502 repositories on GitHub, we provided a quantitative answer to our main research question ($RQ_M$), showing that DDD has matured from a largely theoretical design proposal into an established practical paradigm for engineered software systems in open source.

Our longitudinal analysis identifies 2017 as the critical inflection point where industrial adoption began to accelerate, with project creation and technical intensity reaching sustained peaks in the following years. A key finding of this research is the professional-grade maturity of the verified ecosystem; with an average project longevity of 660.73 days (median: 340.37 days) and a significant organizational footprint of 23.02\%, the DDD landscape is composed of stable, long-lived software that far exceeds the typical activity cycles of the broader GitHub ``long tail''.

Architecturally, the landscape is defined by a shift toward distributed complexity. While traditional Layered Architecture remains prevalent (28.9\%), Clean Architecture (22.78\%) and high-fidelity implementations of CQRS and Event Sourcing have emerged as the principal approaches for managing modern data-intensive systems. Technologically, our data reveals a significant departure from the Java-centric tradition of academic literature, with C\# (34.17\%) and TypeScript (17.71\%) leading industrial adoption. 

Methodologically, our work addresses the ``perils'' of label noise and reproducibility in repository mining. By employing a semantic validation pipeline utilizing LLMs with a triplicate majority-vote strategy, validated against manual expert labelling at substantial agreement (Cohen's $\kappa = 0.77$), we have established a new standard for identifying true ``Rich Domain Models'' at scale. However, we also identify a persistent disconnect between the mining and modeling communities, reflected in the 25.3\% of projects with ``Unknown'' business contexts due to the limitations of version control systems in capturing architectural intent \cite{robles_reflection_2023}.

Future work should focus on closing this gap by developing architectural traceability standards, where strategic design decisions and bounded context boundaries are explicitly tagged within the commit history \cite{robles_reflection_2023}. Additionally, we envision extending our semantic pipeline to resolve ``Unknown'' domains through zero-shot code analysis and performing longitudinal studies on the effects of DDD adoption on architectural decay and technical debt management in multi-stakeholder environments \cite{ozkan_domain-driven_2025}. Ultimately, this study serves as a data-driven baseline, confirming that DDD is no longer solely a theoretical design philosophy but a widely adopted, industrial-grade practice in modern software engineering.

\section*{Declaration of Competing Interest}
The authors declare that they have no known competing financial interests or personal relationships that could have appeared to influence the work reported in this paper.

\section*{Data Availability}
Data will be made available on request.

\bibliographystyle{elsarticle-num-names}
\bibliography{references}

@book{evans_domain-driven_2004,
  title = {Domain-Driven Design: Tackling Complexity in the Heart of Software},
  shorttitle = {Domain-Driven Design},
  author = {Evans, Eric},
  year = {2004},
  publisher = {Addison-Wesley},
  address = {Boston},
  isbn = {978-0-321-12521-7},
  lccn = {QA76.76.D47 E82 2004}
}

@inproceedings{hindle_automated_2011,
  title = {Automated Topic Naming to Support Cross-Project Analysis of Software Maintenance Activities},
  booktitle = {Proceedings of the 8th {{Working Conference}} on {{Mining Software Repositories}}},
  author = {Hindle, Abram and Ernst, Neil A. and Godfrey, Michael W. and Mylopoulos, John},
  year = {2011},
  month = may,
  pages = {163--172},
  publisher = {ACM},
  address = {Waikiki, Honolulu HI USA},
  doi = {10.1145/1985441.1985466},
  urldate = {2024-06-15},
  isbn = {978-1-4503-0574-7},
  langid = {english}
}

@article{kvalseth_note_1989,
  title = {Note on {{Cohen}}'s {{Kappa}}},
  author = {Kv{\aa}lseth, Tarald O.},
  year = {1989},
  month = aug,
  journal = {Psychological Reports},
  volume = {65},
  number = {1},
  pages = {223--226},
  issn = {0033-2941, 1558-691X},
  doi = {10.2466/pr0.1989.65.1.223},
  urldate = {2024-06-15},
  copyright = {http://journals.sagepub.com/page/policies/text-and-data-mining-license},
  langid = {english}
}

@article{liao_prediction_2019,
  title = {A {{Prediction Model}} of the {{Project Life-Span}} in {{Open Source Software Ecosystem}}},
  author = {Liao, Zhifang and Zhao, Benhong and Liu, Shengzong and Jin, Haozhi and He, Dayu and Yang, Liu and Zhang, Yan and Wu, Jinsong},
  year = {2019},
  month = aug,
  journal = {Mobile Networks and Applications},
  volume = {24},
  number = {4},
  pages = {1382--1391},
  issn = {1383-469X, 1572-8153},
  doi = {10.1007/s11036-018-0993-3},
  urldate = {2024-12-07},
  langid = {english}
}

@article{liew_using_2022,
  title = {Using {{Large Language Models}} to {{Generate Engaging Captions}} for {{Data Visualizations}}},
  author = {Liew, Ashley and Mueller, Klaus},
  year = {2022},
  doi = {10.48550/ARXIV.2212.14047},
  urldate = {2024-01-06},
  copyright = {Creative Commons Attribution 4.0 International}
}

@article{ozkan_domain-driven_2025,
  author    = {Ozan {\"O}zkan and {\"O}nder Babur and Mark van den Brand},
  title     = {Domain-Driven Design in Software Development: A Systematic Literature Review on Implementation, Challenges, and Effectiveness},
  journal   = {Journal of Systems and Software},
  volume    = {230},
  pages     = {112537},
  year      = {2025},
  doi       = {10.1016/j.jss.2025.112537}
}

@article{ozkan_refactoring_2023-1,
  title = {Refactoring with Domain-Driven Design in an Industrial Context: {{An}} Action Research Report},
  shorttitle = {Refactoring with Domain-Driven Design in an Industrial Context},
  author = {{\"O}zkan, Ozan and Babur, {\"O}nder and Van Den Brand, Mark},
  year = {2023},
  month = jul,
  journal = {Empirical Software Engineering},
  volume = {28},
  number = {4},
  pages = {94},
  issn = {1382-3256, 1573-7616},
  doi = {10.1007/s10664-023-10310-1},
  urldate = {2024-01-03},
  langid = {english}
}

@inproceedings{panichella_how_2013,
  title = {How to Effectively Use Topic Models for Software Engineering Tasks? {{An}} Approach Based on {{Genetic Algorithms}}},
  shorttitle = {How to Effectively Use Topic Models for Software Engineering Tasks?},
  booktitle = {2013 35th {{International Conference}} on {{Software Engineering}} ({{ICSE}})},
  author = {Panichella, Annibale and Dit, Bogdan and Oliveto, Rocco and Di Penta, Massimilano and Poshynanyk, Denys and De Lucia, Andrea},
  year = {2013},
  month = may,
  pages = {522--531},
  publisher = {IEEE},
  address = {San Francisco, CA, USA},
  doi = {10.1109/ICSE.2013.6606598},
  urldate = {2024-06-15},
  isbn = {978-1-4673-3076-3 978-1-4673-3073-2}
}

@article{ralph_empirical_2020,
  title = {Empirical {{Standards}} for {{Software Engineering Research}}},
  author = {Ralph, Paul and bin Ali, Nauman and Baltes, Sebastian and Bianculli, Domenico and Diaz, Jessica and Dittrich, Yvonne and Ernst, Neil and Felderer, Michael and Feldt, Robert and Filieri, Antonio and {de Fran{\c c}a}, Breno Bernard Nicolau and Furia, Carlo Alberto and Gay, Greg and Gold, Nicolas and Graziotin, Daniel and He, Pinjia and Hoda, Rashina and Juristo, Natalia and Kitchenham, Barbara and Lenarduzzi, Valentina and Mart{\'i}nez, Jorge and Melegati, Jorge and Mendez, Daniel and Menzies, Tim and Molleri, Jefferson and Pfahl, Dietmar and Robbes, Romain and Russo, Daniel and Saarim{\"a}ki, Nyyti and Sarro, Federica and Taibi, Davide and Siegmund, Janet and Spinellis, Diomidis and Staron, Miroslaw and Stol, Klaas and Storey, Margaret-Anne and Taibi, Davide and Tamburri, Damian and Torchiano, Marco and Treude, Christoph and Turhan, Burak and Wang, Xiaofeng and Vegas, Sira},
  year = {2020},
  doi = {10.48550/ARXIV.2010.03525},
  urldate = {2024-01-20},
  copyright = {Creative Commons Attribution Non Commercial Share Alike 4.0 International}
}

@article{spinellis_software_2021,
  title = {Software Evolution: The Lifetime of Fine-Grained Elements},
  shorttitle = {Software Evolution},
  author = {Spinellis, Diomidis and Louridas, Panos and Kechagia, Maria},
  year = {2021},
  month = feb,
  journal = {PeerJ Computer Science},
  volume = {7},
  pages = {e372},
  issn = {2376-5992},
  doi = {10.7717/peerj-cs.372},
  urldate = {2024-12-07},
  copyright = {https://creativecommons.org/licenses/by/4.0/},
  langid = {english}
}

@book{vernon_implementing_2013,
  title = {Implementing Domain-Driven Design},
  author = {Vernon, Vaughn},
  year = {2013},
  publisher = {Addision-Wesley},
  address = {Upper Saddle River, NJ},
  isbn = {978-0-321-83457-7},
  lccn = {QA76.76.D47 V45 2013}
}

@book{gamma_design_2009,
	address = {Reading, MA},
	title = {Design patterns: elements of reusable object-oriented software},
	isbn = {978-0-321-70069-8},
	shorttitle = {Design patterns},
	language = {eng},
	publisher = {Addison-Wesley},
	author = {Gamma, Erich and Helm, Richard and Johnson, Ralph E. and Vlissides, John},
	year = {2009}
}

@book{buschmann_pattern-oriented_1996,
	address = {Chichester ; New York},
	title = {Pattern-oriented software architecture: a system of patterns},
	isbn = {978-0-471-95869-7},
	shorttitle = {Pattern-oriented software architecture},
	publisher = {Wiley},
	editor = {Buschmann, Frank},
	year = {1996},
}

@misc{brandolini_eventstorming_2013,
  author = {Brandolini, Alberto},
  title = {Introducing EventStorming – An Explosive Modeling Technique for Lean Discovery},
  year = {2013},
  howpublished = {\url{https://ziobrando.blogspot.com/2013/11/introducing-eventstorming.html}}
}

@inproceedings{chabert_experienced_2022,
    title = {Mining Experienced Developers in Open-source Projects},
    author = {Chabert, Lucas and Jouve, Pierre-Antoine and D'Hondt, Théo and M\'{e}gret, Olivier and Blanc, Xavier},
    booktitle = {Proceedings of the 17th International Conference on Mining Software Repositories (MSR)},
    year = {2022},
    pages = {23--34},
    doi = {10.1109/MSR55620.2022.00010},
}

@article{maia_systematic_2022,
    title = {A systematic literature review on data cleaning and filtering in mining software repositories},
    author = {Maia, Diogo and Spinu, Cristina and An{\'o}s, Francisco},
    journal = {Empirical Software Engineering},
    volume = {27},
    number = {4},
    pages = {1--53},
    year = {2022}
}

@inproceedings{kochanthara_painting_2022,
  author = {Kochanthara, Sangeeth and Dajsuren, Yanja and Cleophas, Loek and van den Brand, Mark},
  title = {Painting the Landscape of Automotive Software in GitHub},
  booktitle = {Proceedings of the 19th International Conference on Mining Software Repositories},
  series = {MSR '22},
  year = {2022},
  isbn = {978-1-4503-9303-4/22/05},
  location = {Pittsburgh, PA, USA},
  publisher = {Association for Computing Machinery},
  address = {New York, NY, USA},
  doi = {10.1145/3524842.3528460},
  url = {https://doi.org/10.1145/3524842.3528460},
  pages = {1--12}
}

@inproceedings{rigas_mining_2023,
  author = {Rigas, Emmanouil S. and Papoutsoglou, Maria and Kapitsaki, Georgia M. and Bassiliades, Nick},
  title = {Mining Software Repositories to Identify Electric Vehicle Trends: The Case of {GitHub}},
  booktitle = {Proceedings of the 2023 IEEE/ACM International Conference on Software Engineering},
  year = {2023}
}

@article{guemes_emerging_2018,
  author = {G{\"u}emes-Pe{\~n}a, Diego and L{\'o}pez-Nozal, Carlos and Marticorena-S{\'a}nchez, Ra{\'u}l and Maudes-Raedo, Jes{\'u}s},
  title = {Emerging topics in mining software repositories: Machine learning in software repositories and datasets},
  journal = {Progress in Artificial Intelligence},
  year = {2018},
  publisher = {Springer}
}

@article{kalliamvakou_promises_2016,
  author = {Kalliamvakou, Eirini and Gousios, Georgios and Blincoe, Kelly and Singer, Leif and German, Daniel M. and Damian, Daniela},
  title = {An in-depth study of the promises and perils of mining GitHub},
  journal = {Empirical Software Engineering},
  volume = {21},
  number = {5},
  pages = {2035--2071},
  year = {2016},
  doi = {10.1007/s10664-015-9393-5}
}

@article{munaiah_curating_2017,
  title = {Curating GitHub for engineered software projects},
  author = {Munaiah, Nuthan and Kroh, Steven and Cabrey, Craig and Nagappan, Meiyappan},
  journal = {Empirical Software Engineering},
  volume = {22},
  number = {6},
  pages = {3219--3253},
  year = {2017},
  doi = {10.1007/s10664-017-9512-6}
}

@article{cosentino_systematic_2017,
  author = {Cosentino, Valerio and C\'{a}novas Izquierdo, Javier L. and Cabot, Jordi},
  title = {A Systematic Mapping Study of Software Development With GitHub},
  journal = {IEEE Access},
  volume = {5},
  pages = {7173--7192},
  year = {2017},
  doi = {10.1109/ACCESS.2017.2682323}
}

@article{nikoo_empirical_2025,
  title = {An empirical study of business process models and model clones on GitHub},
  author = {Saeedi Nikoo, Mahdi and Kochanthara, Sangeeth and Babur, \"{O}nder and van den Brand, Mark},
  journal = {Empirical Software Engineering},
  volume = {30},
  number = {48},
  year = {2025},
  doi = {10.1007/s10664-024-10584-z}
}

@article{almarzouq_mining_2020,
  author = {AlMarzouq, Mohammad and AlZaidan, Abdullatif and AlDallal, Jehad},
  title = {Mining GitHub for research and education: challenges and opportunities},
  journal = {International Journal of Web Information Systems},
  volume = {16},
  number = {4},
  pages = {451--473},
  year = {2020},
  doi = {10.1108/IJWIS-03-2020-0016}
}

@article{robles_reflection_2023,
  title = {A reflection on the impact of model mining from {GitHub}},
  author = {Robles, Gregorio and Chaudron, Michel R. V. and Jolak, Rodi and Hebig, Regina},
  journal = {Information and Software Technology},
  volume = {164},
  pages = {107317},
  year = {2023},
  doi = {10.1016/j.infsof.2023.107317}
}

@article{tutko_how_2022,
  title = {How are {Software} {Repositories} {Mined}? {A} {Systematic} {Literature} {Review} of {Workflows}, {Methodologies}, {Reproducibility}, and {Tools}},
  author = {Tutko, Adam and Mockus, Audris and Henley, Austin Z.},
  journal = {arXiv preprint arXiv:2204.08108},
  year = {2022},
  url = {https://arxiv.org/abs/2204.08108}
}

@book{wohlin_experimentation_2012,
  title = {Experimentation in {Software} {Engineering}},
  author = {Wohlin, Claes and Runeson, Per and H{\"o}st, Martin and Ohlsson, Magnus C. and Regnell, Bj{\"o}rn and Wessl{\'e}n, Anders},
  publisher = {Springer Science \& Business Media},
  year = {2012},
  isbn = {978-3-642-29043-5}
}

@article{hou_large_2023,
  title     = {Large Language Models for Software Engineering: A Systematic Literature Review},
  author    = {Hou, Xinyi and Zhao, Yanjie and Liu, Yue and Yang, Zhou and Wang, Kailong and Li, Li and Luo, Xiapu and Lo, David and Grundy, John and Wang, Haoyu},
  journal   = {ACM Transactions on Software Engineering and Methodology},
  volume    = {33},
  number    = {8},
  articleno = {220},
  pages     = {1--79},
  year      = {2024},
  publisher = {ACM},
  doi       = {10.1145/3695988}
}

@misc{github_octoverse_2020,
  author = {{GitHub}},
  title  = {The State of the Octoverse 2020},
  year   = {2020},
  howpublished = {\url{https://octoverse.github.com/2020/}}
}

@article{landis_measurement_1977,
  title = {The Measurement of Observer Agreement for Categorical Data},
  author = {Landis, J. Richard and Koch, Gary G.},
  year = {1977},
  journal = {Biometrics},
  volume = {33},
  number = {1},
  pages = {159--174},
  publisher = {International Biometric Society},
  issn = {0006-341X},
  doi = {10.2307/2529310}
}

@misc{msci_gics_2024,
  author       = {{MSCI and S\&P Dow Jones Indices}},
  title        = {Global Industry Classification Standard ({GICS}) Methodology},
  year         = {2024},
  howpublished = {\url{https://www.msci.com/indexes/index-resources/gics}},
  note         = {Industry classification taxonomy originally developed in 1999}
}

@misc{refinitiv_trbc_2020,
  author       = {{Refinitiv (London Stock Exchange Group)}},
  title        = {The Refinitiv Business Classification ({TRBC}) Methodology},
  year         = {2020},
  howpublished = {\url{https://www.lseg.com/en/data-analytics/financial-data/indices/trbc-business-classification}},
  note         = {Originally the Reuters Business Sector Scheme (2004)}
}

@inproceedings{maddodi_aggregate_2020,
  author    = {Maddodi, Gururaj and Jansen, Slinger and Overeem, Michiel},
  title     = {Aggregate Architecture Simulation in Event-Sourcing Applications using Layered Queuing Networks},
  booktitle = {Proceedings of the ACM/SPEC International Conference on Performance Engineering (ICPE '20)},
  pages     = {238--245},
  year      = {2020},
  publisher = {ACM},
  doi       = {10.1145/3358960.3375797}
}

@article{hippchen_designing_2017,
  author  = {Hippchen, Benjamin and Giessler, Pascal and Steinegger, Roland H. and Schneider, Michael and Abeck, Sebastian},
  title   = {Designing Microservice-Based Applications by Using a Domain-Driven Design Approach},
  journal = {International Journal on Advances in Software},
  volume  = {10},
  number  = {3\&4},
  pages   = {432--445},
  year    = {2017}
}

@inproceedings{krause_microservice_2020,
  author    = {Krause, A. and Zirkelbach, C. and Hasselbring, W. and Lenga, S. and Kr{\"o}ger, D.},
  title     = {Microservice Decomposition via Static and Dynamic Analysis of the Monolith},
  booktitle = {2020 IEEE International Conference on Software Architecture Companion (ICSA-C)},
  pages     = {9--16},
  year      = {2020},
  publisher = {IEEE},
  doi       = {10.1109/ICSA-C50368.2020.00011}
}

@inproceedings{joselyne_partitioning_2018,
  author    = {Jos{\'e}lyne, M. I. and Tuheirwe-Mukasa, D. and Kanagwa, B. and Balikuddembe, J.},
  title     = {Partitioning Microservices: A Domain Engineering Approach},
  booktitle = {Proceedings of the International Conference on Software Engineering in Africa (SEiA)},
  pages     = {43--49},
  year      = {2018},
  publisher = {ACM}
}

@inproceedings{kapferer_domain-driven_2020,
  author    = {Kapferer, Stefan and Zimmermann, Olaf},
  title     = {Domain-Driven Service Design: Context Modeling, Model Refactoring and Contract Generation},
  booktitle = {Service-Oriented Computing (SummerSoC 2020)},
  series    = {Communications in Computer and Information Science},
  volume    = {1310},
  pages     = {189--208},
  year      = {2020},
  publisher = {Springer},
  doi       = {10.1007/978-3-030-64846-6\_11}
}

@inproceedings{ait_survival_2022,
  author    = {Ait, Adem and C{\'a}novas Izquierdo, Javier Luis and Cabot, Jordi},
  title     = {An Empirical Study on the Survival Rate of GitHub Projects},
  booktitle = {Proceedings of the 19th International Conference on Mining Software Repositories (MSR)},
  pages     = {365--375},
  year      = {2022},
  publisher = {IEEE/ACM},
  doi       = {10.1145/3524842.3527941}
}

@inproceedings{uludag_supporting_2018,
  author    = {Uluda{\u{g}}, {\"O}mer and Kleehaus, Martin and Matthes, Florian},
  title     = {Supporting Large-Scale Agile Development with Domain-Driven Design},
  booktitle = {Agile Processes in Software Engineering and Extreme Programming (XP 2018)},
  series    = {Lecture Notes in Business Information Processing},
  volume    = {314},
  year      = {2018},
  publisher = {Springer},
  doi       = {10.1007/978-3-319-91602-6\_16}
}

@inproceedings{hebig_quest_2016,
  author    = {Hebig, Regina and Ho-Quang, Truong and Chaudron, Michel R. V. and Robles, Gregorio and Fern{\'a}ndez, Miguel A.},
  title     = {The Quest for Open Source Projects that Use {UML}: Mining {GitHub}},
  booktitle = {Proceedings of the ACM/IEEE 19th International Conference on Model Driven Engineering Languages and Systems (MODELS)},
  pages     = {173--183},
  year      = {2016},
  publisher = {ACM},
  doi       = {10.1145/2976767.2976778}
}

@misc{github_about_2025,
  author       = {{GitHub}},
  title        = {About GitHub},
  howpublished = {\url{https://github.com/about}},
  year         = {2025},
  note         = {Accessed 2025}
}

@misc{tiobe_index_2026,
  author       = {{TIOBE Software BV}},
  title        = {{TIOBE} Index},
  year         = {2026},
  howpublished = {\url{https://www.tiobe.com/tiobe-index/}},
  note         = {Accessed: 2026-07-07}
}




\end{document}